\documentclass{elsart5p}

\usepackage{algorithm}
\usepackage{algpseudocode}

\usepackage{amsmath}
\usepackage{amssymb}
\usepackage{graphicx}

\graphicspath{{graphics/}}

\newcommand{\eat}[1]{}
\newcommand{\sep}{{\bf t\!\!-\!\!sep}}
\begin{document}

\begin{frontmatter}



\title{Local and Global Analysis of Parametric Solid Sweeps}


 \author{Bharat Adsul, Jinesh Machchhar, Milind Sohoni}          % Important note: Anonymize your paper for double-blind reviewing!
                                     % Important note: Anonymize your paper for double-blind reviewing!
%\address{Anonymous Author's Address} % Important note: Anonymize your paper for double-blind reviewing!

\begin{abstract}
In this work, we propose a detailed computational framework for modelling the
envelope of the swept volume, that is the boundary of the volume obtained by
sweeping an input solid along a trajectory of rigid motions. Our framework 
is adapted to the well-established
industry-standard brep format to enable its implementation in modern CAD
systems. This is achieved via a ``local analysis'', which covers
parametrization and singularities, as well as a ``global theory'' which
tackles face-boundaries, self-intersections and trim curves.  Central to
the local analysis is the ``funnel'' which serves as a natural parameter space
for the basic surfaces constituting the sweep. The trimming problem is
reduced to the problem of surface-surface intersections of these basic surfaces.
Based on the complexity of these intersections, we introduce a
novel classification of sweeps as either decomposable or non-decomposable. Further, 
we construct an {\em invariant} function $\theta$ on the funnel which 
\eat{allows us to} 
efficiently separates decomposable and non-decomposable sweeps. 
Through a geometric theorem we also show intimate
connections between $\theta$, local curvatures and
the inverse trajectory used in earlier works as an approach towards trimming.
In contrast to the inverse trajectory approach, $\theta$ is robust
and is the key to a complete structural understanding, and 
\eat{allows} an efficient computation of both, the singular locus and 
the trim curves, which are central to a stable implementation.  
Several illustrative outputs of a pilot implementation are included.
 
\end{abstract}

\begin{keyword}
Sweeping, boundary representation, parametric curves and surfaces
\end{keyword}
\end{frontmatter}

% main text
\section{Introduction} \label{introSec}
This paper is motivated by the need for a robust implementation of solid sweeps in
solid modeling kernels. The solid sweep is of course, the envelope
surface of a solid which is swept in space by a family of rotations and translations.
The uses of sweeps are many, e.g., in the design of scrolls~\cite{scroll}, 
in CNC machining verification~\cite{completeSweep}, 
to detect collisions, and so on. 
See Appendix for an application of solid sweep in designing scrolls,   
where we describe a modeling attempt using an existing kernel and its limitations.
Constant radius blends can be considered as the partial envelope of a sphere 
moving along a specified path.  As with blends, it is expected 
that a deeper mathematical understanding of solid sweep will lead to its rapid 
deployment and use.

A robust implementation of solid sweep poses the following
requirements: (i) allow for input models specified in the
industry-standard brep format, (ii) output the sweep envelope in the brep
format, with effective evaluators,
and finally, (iii) perform body-check, i.e., a check on the orientability,
non-self-intersection, detection of singularities and so on.
Thus there are some ``local'' parts and some ``global'' parts to the
problem.

It is
generally recognized that the harder parts of the local theory is in the
smooth case,
i.e., when faces meet each other smoothly.
For in the non-smooth case, the added complexity in the 
local geometry of the sweep is exactly that of a curve moving in 3-space.
This
is of course well understood, and offered by many kernels as a basic surface
type. As far as we know, the global situation in the non-smooth case, i.e.,
the topological structure of edges and
vertices (i.e., the 1-cage) of the sweep has not been elucidated, but is
also generally assumed
to be simpler than the smooth case.
In fact, much of existing literature has focused on a smooth single-face
solid, as the key problem~\cite{jacobian, sede, trimming}.

In this paper, we focus on the smooth multi-face
solid. In Section~\ref{simpleSec}, we start with the mathematical
structure of the simple sweep (i.e., one without singularities and
self-intersections). By the calculus of curves of contact, we set up a
correspondence between the faces, edges and vertices of
the envelope with those of the swept solid. This sets up the brep structure
of the envelope. Next, we define the funnel as the parametrization space
of a face of the envelope and construct a parametrization. We further elucidate
the structure of the bounding edges/vertices of a face and provide several examples 
of simple sweeps from a pilot implementation.

In Section~\ref{simpleSISec}, we examine the trim structures. The funnel 
of Section~\ref{simpleSec} will remain the ambient parametrization of the faces. 
The correspondence will help us define the trim areas and trim curves which 
must be excised to form the correct envelope. We then define the function $\ell$ 
and use it to define elementary and singular trim curves.

In Section~\ref{decompSec}, we start with the decomposable sweep, i.e., one
which may be partitioned into
a suitable small collection of simple sweeps. The final envelope is obtained
by stable (transversal) boolean operations on this collection. We show that
the trim curves so obtained are elementary. We next define an invariant
$\theta $ on the funnels, which is robustly and efficiently computable
and we show that $\theta >0$ on (all) the funnels characterizes decomposability.
This is an important step in the robust implementation of sweeps.

In Section~\ref{thetaSec}, we prove some of the properties of $\theta $ such as 
its invariance and show that it is the determinant of the transformation
connecting two 2-frames on the envelope, and is thus an easily computable 
function on the surface. We show that the $\theta=0$
curve on the funnel is also the singular locus for the envelope surface.
Via a geometric theorem, we also show that the function $\theta $ matches
the one by~\cite{trimming}  for implicitly defined surfaces and using the so-called
inverse trajectory.

In Section~\ref{nonDecompSec}, we define the singular trim curve, i.e., where $\ell $ may
hit zero. We show that there is a correspondence between singular trim
curves and the curves in the zero-locus of $\theta $. We also show
that (i) singular trim curves make contact with the $\theta=0$ curves, and
(ii) excision at the singular trim curves removes all singularities
of the envelope except at these points of contact. Furthermore, these 
points are easily and robustly computed.

In Section~\ref{conclusionSec} we summarize what has been achieved, viz., that the
decomposability and the zero-locus of $\theta $ complement to give a complete
understanding of all trim curves. We also discuss some implementation issues
and extensions.

\noindent{\bf Previous work}

We now review existing related work. 
Perhaps the most elaborate proposal for the sweep surface ${\cal E}$ is the 
sweep envelope differential equations~\cite{sede} approach, where the authors 
(i) assume that surface $S$ being swept is implicitly given by a function $f$, and (ii) derive a 
differential equation whose solution 
is the envelope.  For any point $p$ on the initial curve of contact, 
a Runge-Kutta marching yields a trajectory  $p(t)$ such that (i) $p(0)=p$, 
and (ii) $p(t) \in {C}(t)$, the curve of contact at time $t$.  
These trajectories presumably serve as the iso-parametric lines $p(t)={\cal E}(t,u(p))$.  
Determining whether $p(t)$ is in the trimming set $T$ is solved by using the inverse trajectory 
condition. This is implemented by using 
the second derivative of the function $\phi(x,t) = f(\eta(x,t))$, where $\eta$ is the inverse trajectory of point $x$. 

On the global front, the building of the envelope ${\cal E}$ is 
done by selecting a collection of points on the initial curve of 
contact, developing trajectories, testing for membership in $T$ and then 
using the points which pass to construct an approximation to the envelope. 
The drawbacks are clear. Typically, constructing an $f$ which defines $S$ is 
difficult. Furthermore, the choice of $f$ seems to determine many computational
and parametric issues, which is undesirable. The inverse-trajectory check 
remains poorly conditioned, especially when the second derivative of 
the function $\phi(x,t)$ w.r.t. $t$ is zero. The structure of the envelope is unknown where 
this derivative is zero.  
A global understanding of $T$ and the nature of
the trim curves is missing.  

In~\cite{classifyPoints}, while classifying points for sweeping solids, the authors give a membership test for a point in the object space to belong inside, outside or on the boundary of the 
swept volume by using inverse trajectory of that point.  
A curve-solid intersection is required to be computed for each point membership query which is computationally expensive, especially when the intersection is non-transversal, as noted by the authors themselves.  Such high degree of computational complexity is prohibitive for a practical implementation.  

In~\cite{planarSwep} the authors work with 2D shapes and 2D motions and quantify singularities using inverse 
trajectories.  This work is based on the computational framework  described in~\cite{classifyPoints}  and involves computing intersections between 2D curves and 2D shapes. 
The authors remark that this work can be extended to the 3-dimensional case involving intersections between 3D curves and 3D solids.  This approach has the same drawback 
as~\cite{classifyPoints}, namely a high computational cost.

In trimming self-intersections in swept volumes~\cite{selfIntersections}, the authors detect self-intersections by computing approximate curves of contact at a few discrete time instances 
which are then checked for intersections.  Approximations are introduced at multiple levels, hence an accurate solution cannot be expected from this method.
%lit survey end
\section{Mathematical structure of sweeps} \label{simpleSec}

In this section we formulate the boundary of the volume obtained by sweeping a solid $M$ along 
a given trajectory $h$.  

\subsection{Correspondence and brep structure of envelope}

\begin{figure}
 \centering
 \includegraphics[scale=0.5]{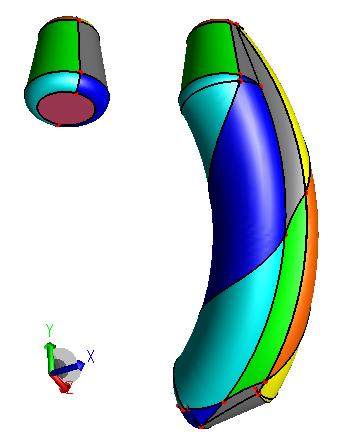}
 \caption{The envelope of a blended cone being swept along a helical trajectory with compounded rotation.}
 \label{coneFig}
\end{figure}

We will use the boundary representation, also known as brep, which is a popular standard for 
representing a compact and oriented solid $M$ by its boundary $\partial M$. The boundary $\partial M$ 
separates the interior of $M$ from the exterior of $M$ and is represented using a set 
of \emph{faces}, \emph{edges} and \emph{vertices}.  See Figure~\ref{coneFig} for the brep of
a solid where different faces are colored differently.  Faces meet in edges and edges meet in vertices.  
The brep consists of two interconnected pieces of information, viz., 
the geometric and the topological.  The geometric information consists of the parametric description of 
the faces and edges while the topological information consists of orientation of the geometric entities and 
adjacency relations between them.

In this paper we consider solids whose boundary is formed by faces meeting smoothly.  In the case 
when the faces do not meet smoothly,  the added complexity in the 
local geometry of the sweep is exactly that of a curve moving in 3-space.
This is of course well understood, and offered by many kernels as a basic 
surface type.  The global geometry and topology for this case will be described in a later paper.

\begin{defn} \label{trajectoryDef}
A {\bf trajectory} in $\mathbb{R}^3$ is specified by a map 
\begin{align*}
h:I \rightarrow (SO(3), \mathbb{R}^3), h(t) = (A(t), b(t))
\end{align*}
where $I$ is a closed interval of $\mathbb{R}$, $ A(t) \in SO(3) \footnote{$SO(3)=\{X \mbox{ is a 3 $\times$3 real matrix} |X^t \cdot X = I, det(X)=1  \}$ is the special orthogonal 
group, i.e. the group of rotational transforms.}, b(t) \in \mathbb{R}^3$.    The parameter $t$ represents time.    
\end{defn}

We assume that $h$ is of class $C^k$ for some $k \geq 2$, i.e., partial derivatives of order up to $k$ exist and are continuous.  

We make the following key assumption about $(M,h)$.

\begin{assum} \label{genericAssum}
The tuple $(M,h)$ is in a {\em general position}.
\end{assum}

\begin{defn}  \label{envlDef}
The {\bf action} of $h$ (at time $t$ in $I$) on $M$ is given 
by $M(t) = \{ A(t) \cdot x + b(t) | x \in M\}$.  
The {\bf swept volume} ${\cal V}$ is the union 
$\displaystyle \bigcup_{t \in I} M(t)$ and the {\bf envelope} ${\cal E}$ is defined as the 
boundary of the swept volume ${\cal V}$.  
\end{defn}

Clearly, for each point $y$ of ${\cal E}$ there must be an $x \in M$ and a $t \in I$ 
such that $y = A(t) \cdot x + b(t)$.  This sets up the following correspondence relation.

\begin{defn} \label{corrDef}
The {\bf correspondence} $R$ is the set of tuples  
$$R=\{(y,x,t) \in {\cal E} \times M \times I| y = A(t) \cdot x + b(t) \}$$ 
For $t_0 \in I$, we set $R_{t_0} := \{(y,x,t) \in R | t = t_0  \}$. 
Similarly, for $y_0 \in {\cal E}$, we define ${_{y_0}} R := \{(y,x,t) \in R |  y = y_0 \}$.
\end{defn}

We will denote the interior of a set $W$ by $W^o$.
It is clear that ${\cal V}^o = \cup_{t \in I} M(t)^o$. Therefore, we have

\begin{lem} \label{intLem}
If $x \in M^o$, then for all $t \in I$,  $A(t) \cdot x + b(t) \notin {\cal E}$.
\end{lem}

Thus, the points in interior of $M$ do not contribute to ${\cal E}$ at all and 
$R \subset {\cal E} \times \partial M \times I$.  This sets up the brep 
structure for ${\cal E}$.
In the sweep example shown in Figure~\ref{coneFig}, 
the correspondence $R$ is illustrated via color coding, i.e.,   
for $(y, x, t) \in R$, the points $y$ and $x$ are shown in the same color.
The general position assumption on $(M,h)$ can be formulated as the condition
that the induced brep topology of ${\cal E}$ remains invariant under a small
perturbation of $(M,h)$.

\begin{lem} \label{preImageLem}
Assuming general position of $(M,h)$, for any $y \in {\cal E}$, there are at 
most three distinct tuples $(y, x_i, t_i)$ for $i = 1,2,3$  which belong to ${_y}R$.
\end{lem}
\noindent {\em Proof.} For distinct tuples $(y, x_1, t_1), (y, x_2, t_2) 
\in {_y} R$, it is clear that $t_1 \neq t_2$, for otherwise $x_1 = x_2$.
Therefore $\partial M(t_1)$ and $\partial M(t_2)$ intersect 
at point $y$.  By Assumption~\ref{genericAssum} 
this intersection is transversal.  Further, by the same assumption, at most $3$ surfaces may intersect in a 
point. \hfill $\square$

\begin{defn} \label{trajXDef}
For a point $x \in M$, define the {\bf trajectory of} $\pmb{x}$ as the map $\gamma_x : I \to \mathbb{R}^3$ 
given by $\gamma_x(t) = A(t) \cdot x + b(t)$ and the velocity $v_x(t)$ as 
$v_x(t) = \gamma_x'(t) = A'(t) \cdot x + b'(t)$.
\end{defn}

For a point $x \in \partial M$, let $N(x)$ be the unit outward normal to $M$ at $x$.  Define the function 
$g: \partial M \times I \to \mathbb{R}$ as 
\begin{align} \label{gEq}
g(x, t) = \left < A(t) \cdot N(x) , v_x(t) \right >
\end{align} 
Thus, $g(x, t)$ is the dot product of the velocity vector with the unit normal at the point $\gamma_x(t) \in \partial M(t)$.

Proposition~\ref{gLem} gives a necessary condition for a point $x \in \partial M$ to 
contribute a point on ${\cal E}$ at time $t$, namely, $\gamma_x(t)$, 
and is a rewording in our notation of the statement in~\cite{sede} that 
{\em the candidate set is the union of the ingress, the egress and the 
grazing set of points}.

\begin{prop} \label{gLem}
For $(y, x, t) \in R$ and $I = [t_0, t_1]$, either 
(i) $g(x,t) = 0$ or 
(ii) $t = t_0$ and $g(x,t) \leq 0$, or 
(iii) $t = t_1$ and $g(x,t) \geq 0$.
\end{prop}
For proof, refer the Appendix.
\begin{defn} \label{cocDef}
For a fixed time instant $t_0 \in I$, the set $\{ \gamma_x(t_0)| x \in \partial M, g(x,t_0) = 0 \}$ is
referred to as the {\bf curve of contact} at $t_0$ and denoted by $C_I(t_0)$. 
Observe that $C_I(t_0) \subset \partial M(t_0)$. The union of the 
curves of contact is referred to as the {\bf contact set} and denoted by $C_I$, i.e., 
$C_I = \displaystyle \bigcup_{t \in I} C_I(t)$.
\end{defn}

In the sweep example in Figure~\ref{dumbbellFig}, the curve of contact at $t=0$ is shown imprinted 
on the solid in red. The curves of contact are referred to as the 
{\em characteristic curves} in~\cite{peternell}.

\begin{defn} \label{projDef}
Define projections $\tau: R \to I$ 
and $Y: R \to {\cal E}$ as:
$\tau(y,x,t) = t \mbox{~~and~~} Y(y,x,t) = y$.
\end{defn}

\begin{defn} \label{simpleDef}
A sweep $(M,h,I)$ is said to be {\bf simple} if for all $t \in I^o$, 
$C_I(t) = Y(R_t)$. \end{defn} 

Note that, by Proposition~\ref{gLem}, for any sweep, we have 
$Y(R_t) \subseteq C_I(t)$. In a simple sweep, we require that $C_I(t) = Y(R_t)$.
In other words, every point on the contact-set appears on the envelope, and 
thus, no {\em trimming} of the contact-set is needed in order to obtain the envelope.

\begin{lem} \label{simpleLem}
For a simple sweep, for all $y \in {\cal E}$, ${_y} R$ is a singleton set.
\end{lem}
\noindent {\em Proof.} We first show that for a simple sweep, for 
$t \neq t'$, $C_I(t) \cap C_I(t') = \emptyset$.  Suppose that $y \in C_I(t) \cap C_I(t')$.
Clearly, $C_I(t) \subset \partial M(t)$ and $C_I(t') \subset \partial M(t')$. 
Hence $y \in \partial M(t) \cap \partial M(t')$.  Since $\partial M(t)$ and 
$\partial M(t')$ intersect transversally, $C_I(t) \cap M^o(t') \neq \emptyset$ and 
$C_I(t') \cap M^o(t) \neq \emptyset$.  It follows by Lemma~\ref{intLem} that 
$C_I(t) \not \subset Y(R_t)$ and $C_I(t') \not \subset Y(R_{t'})$ which
contradicts the fact that $(M,h,I)$ is simple.

Now suppose that there are $2$ tuples $(y, x_i, t_i) \in {_y}R$ for $i= 1,2$.  Since $\partial M$ is free from self-intersections 
it follows that $t_1 \neq t_2$ and $y \in C_I(t_1) \cap C_I(t_2)$ which is a contradiction to the fact that 
$(M,h,I)$ is simple.
\hfill $\square$

\subsection{Parametrizations}	\label{paramSec}
Now we describe parametrizations of the various entities
of the induced brep structure of ${\cal E}$. Here we restrict to the
case of the simple sweep. The more general case is derived from this. 
\eat{As mentioned before, the brep ${\partial M}$ consists
of faces, edges and vertices. These geometric entities give rise to
corresponding entities on ${\cal E}$ (see Figure~\ref{coneFig}).}

\subsubsection{Geometry of faces of ${\cal E}$}\label{keynotation}
Let $F$ be a face of ${\partial M}$. In general, $F$ gives rise to 
multiple faces of ${\cal E}$. Below we describe a natural parametrization of
these faces using the parametrization of the surface underlying the
face $F$.

\begin{defn} \label{parSurfDef}
A {\bf smooth/regular parametric surface} in $\mathbb{R}^3$ is a smooth map $S: \mathbb{R}^2 \to \mathbb{R}^3$
such that at all $(u_0,v_0) \in \mathbb{R}^2$   
$\frac{\partial S}{\partial u}|_{(u_0,v_0)} \in \mathbb{R}^3$ and $\frac{\partial S}{\partial v}|_{(u_0,v_0)} \in \mathbb{R}^3$ are linearly independent.  Here $u$ and $v$ are called 
the parameters of the surface.
\end{defn}

Let $S(u,v)$ be the surface underlying the face $F$ of $\partial M$.

\begin{defn} \label{fDef}
Define the function $f: \mathbb{R}^2 \times I \to \mathbb{R}$ as 
$f(u,v,t) = g(S(u,v), t)$.
\end{defn}
The domain of function $f$ will be referred to as the parameter space.  Note that $f$ is easily 
and robustly computed.
\begin{defn} \label{LFRDef}
For an interval $I = [t_0, t_1]$, we define the following subsets of the parameter space
\begin{align*}
{\cal L} &= \{ (u,v,t_0) \in \mathbb{R}^2 \times \{t_0\} \mbox{ such that } f(u,v,t_0)\leq 0 \} \\
{\cal F} &= \{ (u,v,t) \in \mathbb{R}^2 \times I \mbox{ such that } f(u,v,t)=0 \} \\
{\cal R} &= \{ (u,v,t_1) \in \mathbb{R}^2 \times \{t_1\} \mbox{ such that } f(u,v,t_1)\geq 0 \} 
\end{align*}
\end{defn}
The set ${\cal F}$ will be referred to as the {\bf funnel}.

\begin{figure}
 \centering
 \includegraphics[scale=0.7]{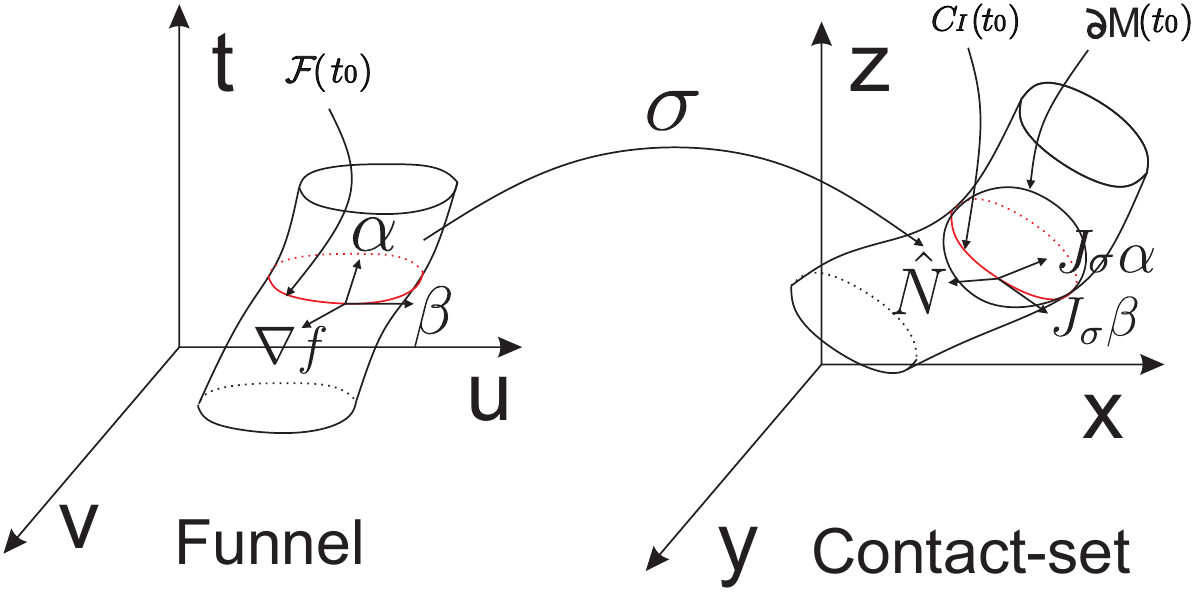}
 \caption{The funnel and the contact-set.}
 \label{funnelFig}
\end{figure}

By Assumption~\ref{genericAssum} about the general position of $(M,h)$ it follows that for all $p \in {\cal F}$, the gradient 
$\nabla f(p) = [f_u (p), f_v(p), f_t(p)]^T \neq \bar{0}$.  As a consequence, ${\cal F}$ is a smooth, orientable surface in the parameter space.

\begin{defn} \label{pcocDef}
The set $\{ (u,v,t) \in {\cal F} | t = t_0 \}$ will be referred to as the {\bf p-curve of contact} at $t_0$ and 
denoted by ${\cal F}(t_0)$.
\end{defn}

We now define the sweep map from the parameter space to the object space.
\begin{defn} \label{sigmaDef}
The {\bf sweep map} is defined as follows.
\begin{align*}
\sigma: \mathbb{R}^2 \times I \to \mathbb{R}^3, \sigma(u,v,t) = A(t) \cdot S(u,v) + b(t)
\end{align*} 
\end{defn}
Note that, $\sigma$ is a smooth map, $C_I = \sigma({\cal F})$ and $C_I(t) = \sigma({\cal F}(t))$. 
Here and later, by a slight abuse of notation, ${\cal E}$, $C_I$ and $C_I(t)$ denote the 
appropriate parts of complete ${\cal E}$, $C_I$ and $C_I(t)$ respectively 
resulting from the face $F \subset \partial M$ whose underlying surface is $S$.
The surface patches $\sigma({\cal L})$ and $\sigma({\cal R})$ will be referred to as the 
left and right end-caps respectively.

The funnel, the contact-set, ${\cal F}(t_0)$ and $C_I(t_0)$ are shown schematically in Figure~\ref{funnelFig}.

The condition $f = 0$ can also be looked upon as the rank deficiency condition~\cite{jacobian} of the 
Jacobian $J_{\sigma}$ of the sweep map $\sigma$.  To make this precise, let
\begin{align} \label{jacobianEq}
J_{\sigma} = \begin{bmatrix} \sigma_u & \sigma_v & \sigma_t \end{bmatrix} _{3 \times 3}
\end{align}
where $\sigma_u = A(t) \cdot \frac{\partial S}{\partial u}(u,v)$, 
$\sigma_v = A(t) \cdot \frac{\partial S}{\partial v}(u,v)$  and 
$\sigma_t = A'(t)\cdot S(u,v) + b'(t)$. Note that if $S(u,v) = x$ then 
$\sigma_t = \gamma_x'(t)$ is the velocity, also denoted by $V(u,v,t)$.   
Observe that  regularity of $S$ ensures 
that $J_{\sigma}$ has rank at least 2.  Further, it is easy to show that 
$f(u,v,t)$ is a non-zero scalar multiple of the determinant of $J_{\sigma}$.
Therefore, the condition $f=0$ is precisely the rank deficiency 
condition of $J_{\sigma}$.

For a simple sweep, by Proposition~\ref{gLem}, Definition~\ref{simpleDef} and Definition~\ref{LFRDef} it 
follows that 
${\cal E} = \sigma({\cal L} \cup {\cal F} \cup {\cal R})$. 
The surface patches $\sigma({\cal L})$ and $\sigma({\cal R})$ can be obtained 
from $\partial M$ using Proposition~\ref{gLem} and Definition~\ref{LFRDef}.  The {\em trim curve}
 in parameter space for $\sigma({\cal L})$ is given by $f(u,v,t_0) = 0 $ and that for 
$\sigma({\cal R})$ is given by $f(u,v,t_1) = 0$.  

We now come to the parametrization of $\sigma({\cal F})$.  The non-singularity of $f$ makes 
${\cal F}$ an effective parametrization space for $\sigma({\cal F})$.
Since time $t$ is a central parameter of the sweep problem and is important in 
numerous applications, it is useful to have $t$ as one of the parameters of $\sigma({\cal F})$.
For most non-trivial sweeps there is no closed form solution for the parametrization of 
the envelope and we address this problem using the procedural paradigm which is now 
standard in many kernels and is described in the Appendix.
In this approach, a set of evaluators are constructed for the 
curve/surface via numerical procedures  which converge to the solution up to the required tolerance.  
This has the advantage of being computationally efficient as well as accurate.

Clearly, the bounding edges of the multiple faces resulting from the
face $F$ of $\partial M$, are generated by the bounding edges of $F$.

\subsubsection{Geometry of edges of ${\cal E}$}
\begin{figure}
 \centering
 \includegraphics[scale=0.6]{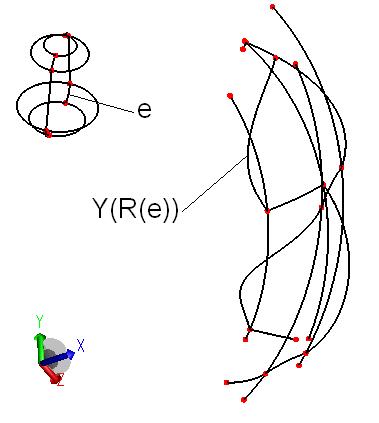}
 \caption{The edges of envelope for the sweep example shown in Figure~\ref{coneFig}.}
 \label{wireFrameFig}
\end{figure}

We now briefly describe the computation of edges of ${\cal E}$.  If $\partial M$ is 
composed of faces meeting smoothly, an edge $e$ of $\partial M$ will, in general, give rise to a 
set of edges in ${\cal E}$.
We define the restriction of $R$ to the edge $e$ as follows.
\begin{defn}
For an edge $e \in \partial M$, define $R(e) = \{ (y,x,t) \in R|x \in e \}$.
\end{defn}

Let $e$ be the intersection of faces $F_1$ and $F_2$ in $\partial M$ and let $s$ denote 
the parameter of $e$.  Since $F_1$ and $F_2$ meet smoothly at $e$, at every point 
$e(s)$ of $e$ there is a well-defined normal.  Hence we may define the following function 
on the parameter space $\mathbb{R} \times I$.

\begin{defn} \label{feDef}
Define the function $f^e: \mathbb{R} \times I \to \mathbb{R}$ as 
$f^e(s,t) = g(e(s),t)$.
\end{defn}

Note that the function $f^e$ is the restriction of the function $f$ defined in Definition~\ref{fDef} to 
the parameter space curve $(u(s), v(s))$ corresponding to the edge $e$ so that $e(s) = S(u(s), v(s))$ 
where $S$ is the surface underlying face $F_1$.  The following Lemma gives a necessary condition for 
a point $e(s)$ to be on ${\cal E}$ at time $t$.
\begin{lem} \label{feLem}
For $(y, e(s), t) \in R(e)$ and $I = [t_0, t_1]$, either (i) $t = t_0$ and $f^e(s,t) \leq 0$, or 
(ii) $t = t_1$ and $f^e(s,t) \geq 0$, or (iii) $f^e(s,t) = 0$.
\end{lem}
\noindent {\em Proof.}  This follows from Prop.~\ref{gLem} and Definition~\ref{feDef}. 
\hfill $\square$
Figure~\ref{wireFrameFig} shows the edges of the envelope for 
the sweep example shown in Figure~\ref{coneFig}.  The correspondence for one of the edges of the envelope is also marked.

Let ${\cal F}_1$ denote the funnel corresponding to the contact set generated by face $F_1$.
The edge in parameter space which bounds ${\cal F}_1$ is given by 
$\{ (u(s), v(s), t) \in \mathbb{R}^2 \times I | f^e(s,t) = 0 \}$ which we will denote by 
${\cal F}^e$.  Note that ${\cal F}^e$ is smooth if 
$(f^e_s, f^e_t) = (f_u \cdot u_s + f_v \cdot v_s, f_t) \neq (0,0)$ at all points
in ${\cal F}^e$.

\subsubsection{Geometry of vertices of ${\cal E}$}

A vertex $z$ on ${\partial M}$ will, in general, give rise to a set of vertices on ${\cal E}$.  We further 
restrict the correspondence $R$ to $z$ as $R(z) = \{(y, x, t) \in R | x = z \}$.  As $\partial M$ is smooth, there is a well-defined normal at $z$.  Hence we may define the function 
$f^z: I \to \mathbb{R}$ as $f^z(t) = g(z,t)$.  If $z$ is on the boundary of a face $F_1$, $z$ will 
have a set of coordinates in the parameter space of the surface $S$ underlying the face $F_1$, say $(u_0, v_0)$, 
so that $z = S(u_0, v_0)$.  It is easy to see that if $(y,z,t) \in R(z)$ and $I = [t_0, t_1]$ then 
either (i) $t = t_0$ and $f^z(t) \leq 0$, or (ii) $t = t_1$ and $f^z(t) \geq 0$, or (iii) $f^z(t) = 0$.

\subsection{Examples of simple sweeps}

\begin{figure}
 \centering
 \includegraphics[scale=0.4]{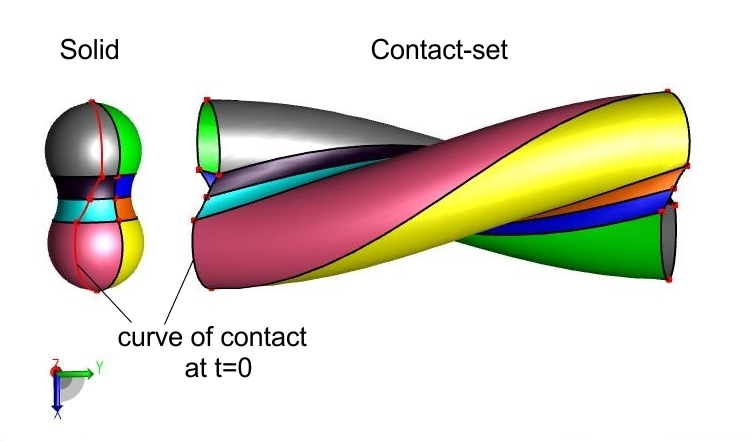}
 \caption{The envelope(without end-caps) of a dumbbell undergoing translation along $y$-axis and undergoing rotation about $y$-axis.}
 \label{dumbbellFig}
\end{figure}

\begin{figure}
 \centering
 \includegraphics[scale=0.5]{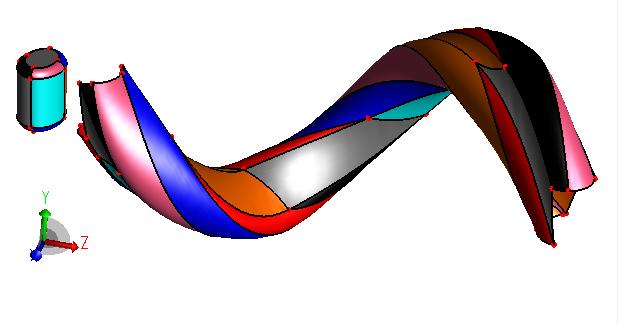}
 \caption{The envelope(without end-caps) of an elliptical cylinder undergoing a screw motion while rotating about 
its own axis.}
 \label{ellipBottleFig}
\end{figure}

\begin{figure}
 \centering
 \includegraphics[scale=0.55]{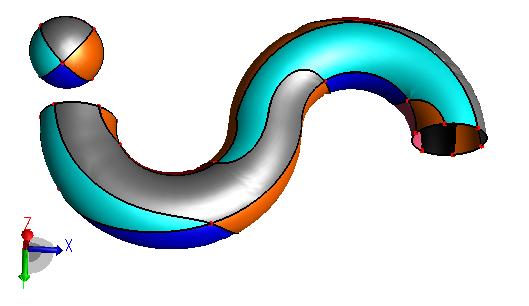}
 \caption{The envelope(without end-caps) of a sphere sweeping along an 'S' shaped trajectory while rotating 
about $y$-axis}
 \label{sphereSFig}
\end{figure}

Three examples of simple sweeps are shown in Figures~\ref{dumbbellFig}, \ref{ellipBottleFig} 
and \ref{sphereSFig} which 
were generated using a pilot implementation of our algorithm in ACIS 3D Modeler \cite{acis}.
A curve of contact at initial time is shown imprinted on the solid in Figure~\ref{dumbbellFig}.

%\section{Simple self-intersections}	\label{simpleSISec}
\section{The trim structures}	\label{simpleSISec}

Unlike in a simple sweep, all points of $C_I$ may not belong to the envelope.  We now define the 
subset of $C_I$ which needs to be excised in order to obtain ${\cal E}$.
\begin{defn} \label{trimSetDef}
The {\bf trim set} is defined as 
$$T_I:=\{ x \in C_I |\exists t \in I, x \in M^o(t) \}$$ 
\end{defn}

\begin{lem} \label{trimSetLem}
The set $T_I$ is open in $C_I$.
\end{lem}
\noindent {\em Proof.}  Consider a point $y_0 \in T_I$.  Then $y_0 \in M^o(t_0)$ for some $t_0 \in I$.  
Hence, there exists an open ball of non-zero radius $r$ 
centered at $y_0$, denote it by $B(y_0, r)$, which is itself contained in $M^o(t_0)$.  
Let ${\cal N}_0 := B(y_0, r) \cap C_I$.
Then, ${\cal N}_0 \subset T_I$ and  ${\cal N}_0$ is open in $T_I$. Hence $T$ is open in $C_I$.
\hfill $\square$

In general, the trim set will span several parts of $C_I$ corresponding 
to different faces of $\partial M$. For the ease of notation
and presentation, in the rest of this paper, we will analyse the corresponding
trim structures on the {\em funnel} of a fixed face $F$ of $\partial M$.
Thanks to the natural parametrizations (cf. subsection~\ref{paramSec}), 
the migration of these trim structures across different funnels is an easy
implementation detail. In view of this, we carry forward the notation 
developed in subsection~\ref{keynotation} through the rest of this paper.

\begin{defn} \label{trimSetDef} 
The pre-image of $T_I$ on the funnel under the map $\sigma$ will be 
referred to as the {\bf p-trim set}, denoted by $pT_I$, i.e., 
$pT_I = \sigma^{-1}(T_I) \cap {\cal F}$.
\end{defn}
An immediate corollary of Lemma~\ref{trimSetLem} is:
$pT_I$ is open in ${\cal F}$.

One can also define similar parametric trim areas on the left and right
caps (cf. ${\cal L}$ and ${\cal R}$ from Definition~\ref{LFRDef}) and their counterparts in the object
space. However, for want of space, we assume here that these trim structures are empty. Our analysis can
be extended to also cover the non-empty case.

\begin{defn} \label{trimCurveDef}
The boundary of $\overline{T_I}$ will be referred to as the {\bf trim curves} and denoted by $\partial T_I$. Here $\overline{T_I}$ denotes the closure of 
$T_I$ in $C_I$. Similarly,
the boundary of the closure $\overline{pT_I}$ of $pT_I$ in ${\cal F}$ 
will be referred to as the {\bf p-trim curves} and denoted by $\partial pT_I$.
\end{defn}

Note that ${\cal E} \cap T_I = \emptyset$,
${\cal E} \cap \overline{T_I} = \partial T_I$ and 
$\sigma({\cal F} \setminus pT_I) = {\cal E}$.
Therefore the problem of excising the trim set is 
reduced to the problem of computing the trim curves. Further, 
this computation is eventually reduced to {\em guided} parametric 
surface-surface intersections via the parametrization of $\sigma({\cal F})$ 
described in subsection~\ref{paramSec}.

\begin{figure}
 \centering
 \includegraphics[scale=0.65]{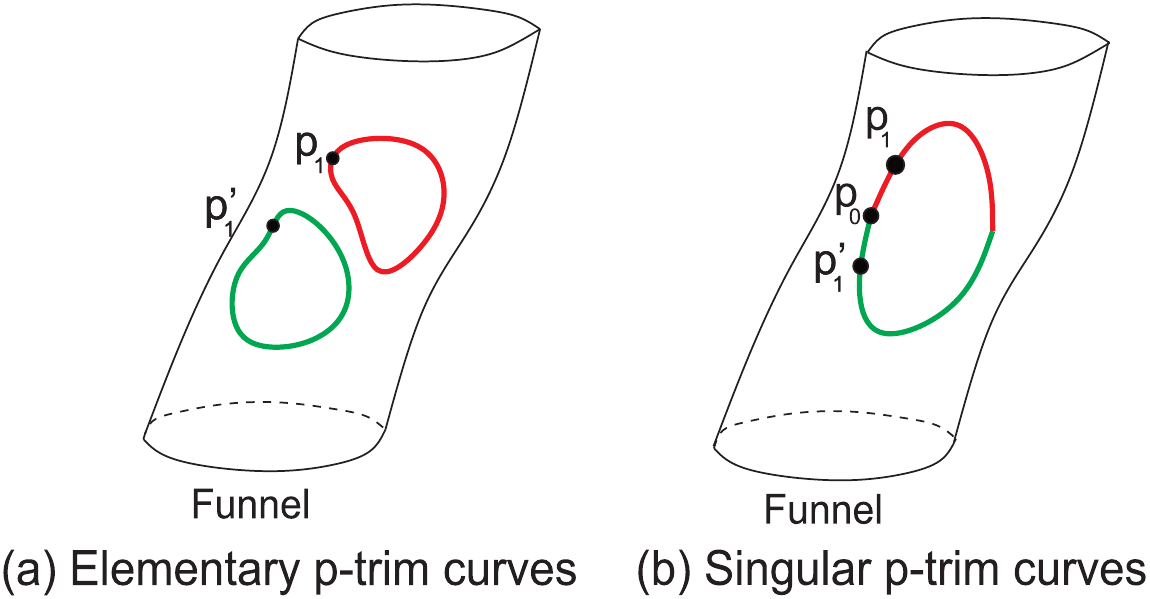}
 \caption{Elementary and singular p-trim curves.}
 \label{ptrimCurveAFig}
\end{figure}

For each point $y \in \partial T_I$ there is a finite set of points $p_i \in \partial pT_I$ such that $\sigma(p_i) = y$ 
for all $i$ (cf. Lemma~\ref{preImageLem}). 
Figure~\ref{ptrimCurveAFig} schematically illustrates p-trim curves on ${\cal F}$.  For every 
point $p_1$ in the red portion of $\partial pT_I$, there is a point $p_1'$ in the green portion of $\partial pT_I$ 
such that $\sigma(p_1) = \sigma(p_1')$.

We extend the correspondence of Definition~\ref{corrDef} 
to $C_I \times M \times I$ as below. Abusing notation, henceforth, $R$ will denote this correspondence. 
\eat{
We extend the correspondence defined in Definition~\ref{corrDef} to $C_I \times M \times I$. By abuse of notation, we denote this extended relation again by 
$R$ and the following definition of $R$ will be used in  the rest of the paper.
}
\begin{defn} 
Let 
$R:= \{(y,x,t) \in C_I \times M \times I| y = A(t) \cdot x + b(t) \}$.
As expected, we define $\tau: R \to I$ 
and $Y: R \to C_I$ as:
$\tau(y,x,t) = t \mbox{~~and~~} Y(y,x,t) = y$.
Further, as before,
$R_{t_0} := \{(y,x,t) \in R | t = t_0  \}$,\!
${_{y_0}}\!R := \{(y,x,t) \in R |  y = y_0 \}$.
\end{defn}

A crucial observation is that, unlike the earlier correspondence, {\em $R \not\subset C_I \times \partial M \times I$}.

\begin{defn} \label{lDef}
For $p=(u,v,t) \in {\cal F}$, let $\sigma(p) = y$. Let $L(p) := \tau({_y}R)$. Define the function 
$\ell: {\cal F} \to \mathbb{R} \cup \infty$ as follows.
\begin{align*}
\ell(p) &= \displaystyle \inf_{t' \in L(p)\setminus\{t\} } \|t - t'\|  &\text{ if } L(p) \neq \{t\} \\
	&= \infty  &\text{ if } L(p) = \{t\}
\end{align*}
Further, we define $\sep = \displaystyle \inf_{p \in{\cal F}} \ell(p)$.
\end{defn}

For $p \in {\cal F}$, $L(p)$ is the set of all time instances $t'$ (except $t$) such 
that some point of $M(t')$ coincides with $\sigma(p)$. 
Further, the function $\ell$ gives the `smallest' 
time $\delta t$ such that some point of $M(t \pm \delta t)$ coincides with $\sigma(p)$.

\begin{lem} \label{trimCLem}
Let $p_0 \in \overline{pT_I}$. Then  $ p_0 \in pT_I$ iff $L(p_0)$ contains an interval, and $p_0 \in \partial pT_I$ iff $L(p_0)$ is a discrete set
of cardinality either two or three.
\end{lem}
\noindent {\em Proof.} Suppose first that $p_0 \in pT_I$.  Let $y_0 := \sigma(p_0)$.  Then $y_0 \in T_I$ and 
$y_0 \in M^o(t_0)$ for some $t_0 \in I$.  Let $B(y_0, r)$ be an open ball of radius $r >0$ centered at 
$y_0$ contained in $M^o(t_0)$.  Assume without loss of generality that $A(t_0) = I$ and $b(t_0) = 0$. 
By continuity of the trajectory $h$ it follows that given $r>0$ there exists $\delta t > 0$ such that 
$\| y_0 - A(t_0+\delta t) \cdot y_0 - b(t_0 +\delta t) \| < r$.  Hence, $y_0 \in M^o(t)$ for all 
$t \in [t_0, t_0 + \delta t]$.  In other words, $[t_0, t_0 + \delta t] \in L(p_0)$.

Conversely, suppose that $L(p_0)$ contains an interval $[t_1, t_2]$, i.e., $y_0 \in M(t)$ for all 
$t \in [t_1, t_2]$.  By Assumption~\ref{genericAssum} about the general position of $(M,h)$ it 
follows that $y_0 \in M^o(t)$ for some $t \in [t_1, t_2]$, i.e., $y_0 \in T_I$ and $p_0 \in pT_I$.
We have shown that for $p_0 \in \overline{pT_I}$, $p_0 \in pT_I$ iff $L(p_0)$ contains an interval.  Hence, $L(p_0)$ is discrete iff $p_0 \in \partial pT_I$. 

As $\partial T_I \subset {\cal E}$, 
by Lemma~\ref{preImageLem}, it follows that 
at all but finitely many points $p \in \partial pT_I$,  
$L(p)$ is of cardinality $2$ and at remaining points it is of cardinality $3$. 
\hfill $\square$

We classify trim curves as follows.
\begin{defn} \label{trimCClassifyDef}
A curve $C$ of $\partial pT_I$ is said to be {\bf elementary} if  there exists $\delta > 0$ such that for all 
$p \in C$, $\ell(p) > \delta$. 
It is said to be {\bf singular} if $\displaystyle \inf_{p \in C} \ell(p) = 0$.
\end{defn}

Figures~\ref{ptrimCurveAFig}(a) and~\ref{ptrimCurveAFig}(b) schematically illustrate elementary and singular p-trim curves on ${\cal F}$ respectively.  
Further observe that, $\sep > 0$ in case (a) and 0 in case (b).

Before proceeding further, we introduce the following notation: for 
$J \subset I$, ${\cal F}(J)=\{(u,v,t) \in {\cal F} \mid t \in J\}$.

\begin{lem} \label{transInterLem}
All but finitely many points of elementary trim curves lie on the 
transversal intersections of two surface 
patches $\sigma({\cal F}(I_i))$ and the remaining points lie on the transversal 
intersection of three surface patches $\sigma({\cal F}(I_i))$ 
where, for $i=1,2,3$, $I_i \subset I$ are subintervals.
\end{lem}
\noindent {\em Proof.}  Suppose that all curves of $\partial pT_I$ are elementary, i.e., $\exists \delta >0$ 
such that for all $p \in \partial pT_I$, $\ell (p) > \delta$.
By Lemma~\ref{trimCLem}, 
all but finitely many points $y \in \partial T_I$ have two points $p_1=(u_1, v_1, t_1)$ and 
$p_2= (u_2, v_2, t_2)$ in $\partial pT_I$ such that 
$\sigma(p_1) = \sigma(p_2) = y$.  Let ${\cal F}_1 := {\cal F}([t_1 - \delta, t_1 + \delta])$ 
and  ${\cal F}_2 := {\cal F}([t_2 - \delta, t_2 + \delta])$.  Then 
$y \in \sigma({\cal F}_1) \cap \sigma({\cal F}_2)$.  From Section~\ref{geomThetaSec} we know that 
$\partial M(t_1)$ and $\partial M(t_2)$ are tangential to $\sigma({\cal F}_1)$ and $\sigma({\cal F}_2)$ 
respectively at $y$.  By Assumption~\ref{genericAssum} about general position of $(M,h)$, $\partial M(t_1)$ and 
$\partial M(t_2)$ intersect transversally at $y$.  Hence, $\sigma({\cal F}_1)$ and $\sigma({\cal F}_2) $ intersect 
transversally at $y$.  

 At most finitely many points $y \in \partial T_I$ have three points $p_1, p_2$ and $p_3$ in $\partial pT_I$ such 
that $\sigma(p_i) = y$.  By an argument similar to above, it can be shown that $y$ lies on the transversal 
intersection of three surface patches $\sigma({\cal F}_i)$ for ${\cal F}_i$ corresponding to appropriate 
subintervals $I_i \subset I$.
\hfill $\square$

\begin{figure}
 \centering
 \includegraphics[scale=0.5]{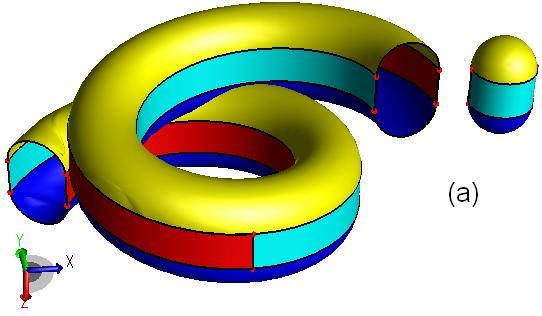}
 \includegraphics[scale=0.5]{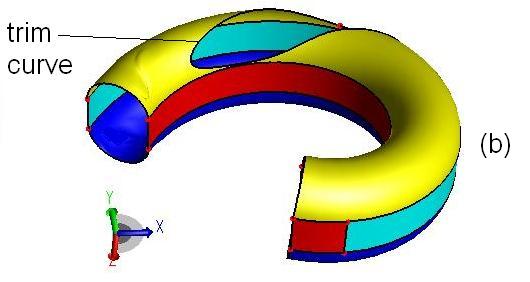}
 \caption{(a) The contact set of a capsule moving along a helix while rotating 
about $y$-axis.(b) The contact set restricted to interval $[0.5,1.0]$ with the trim set excised.}
 \label{toolFig}
\end{figure}

Figure~\ref{toolFig} shows an example in which a capsule is swept along a helical path while rotating about 
$y$-axis.  The trim curves are elementary.

\section{Decomposable sweeps} \label{decompSec}

\begin{figure}
 \centering
 \includegraphics[scale=0.55]{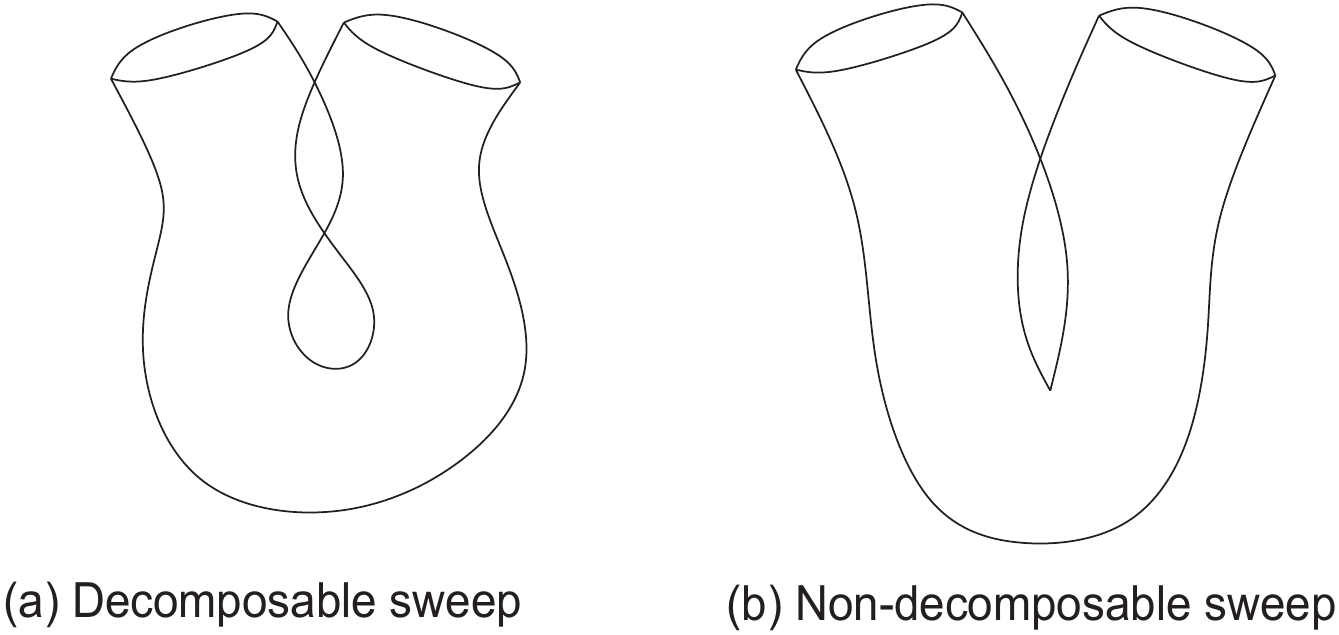}
 \caption{Contact-sets of decomposable and non-decomposable sweeps.}
 \label{decompNonDecompFig}
\end{figure}

We now consider sweeps, which though not simple, can be divided into simple sweeps by 
partitioning the sweep interval so that the trim curves can be obtained by transversal intersections 
of the contact sets of the resulting simple sweeps.
Given an interval $I$, we call a partition ${\cal P}$ of $I$ 
into consecutive intervals $I_1, I_2, \ldots, I_{k_{\cal P}}$ to be of width $\delta$ if $max\{ length(I_1), length(I_2), \ldots, length(I_{k_{\cal P}}) \} = \delta$.
\begin{defn} \label{decompDef}
We say that the sweep $(M,h, I)$ is {\bf decomposable}
if there exists $\delta > 0$ such that for all partitions ${\cal P}$ of $I$ of width $\delta$, 
each sweep $(M,h,I_i )$ is simple for  $i= 1, \cdots, k_{\cal P}$. A sweep
which is not decomposable is called {\bf non-decomposable}.
\end{defn}
Figure~\ref{decompNonDecompFig} schematically illustrates the difference between decomposable 
and non-decomposable sweeps.  The example shown in Figure~\ref{toolFig} is of a decomposable sweep 
in which partitioning the sweep interval $I$ into 2 equal halves will result in 2 simple sweeps. 

\begin{prop} \label{decompLem}
The sweep $(M,h,I)$ is decomposable iff $\sep > 0$. Further, if $\sep > 0$ 
then all the p-trim curves are elementary.
\end{prop}
\noindent {\em Proof.} Suppose first that $\sep > 0$.  Let ${\cal P}$ be a partition of $I$ of width 
$\sep$.  We show that $(M,h,I_i)$ is simple for $i = 1,2,\ldots, k_{\cal P}$. 
Let ${\cal E}_i$ and $C_{I_i}$ be the envelope and the contact set for $(M,h,I_i)$ respectively.
\eat{
and $R^i$ be the correspondence for $(M,h,I_i)$ ,i.e., 
$R^i := \{ (y,x,t) \in  C_{I_i} \times M \times I_i | y = A(t) \cdot x + b(t) \}$. Further, we set $R^i_{t_0} := \{(y,x,t) \in R^i | t = t_0 \}$.}
By Proposition~\ref{gLem}, (modulo end-caps), ${\cal E}_i \subset C_{I_i}$. 
It needs to be shown that $C_{I_i} \subset {\cal E}_i$.  
Suppose not.  Let $y \in C_{I_i}(t)$ such that 
$y \notin {\cal E}_i$ for some $t \in I_i$.  Then, $y \in T_{I_i}$, i.e.,  $y \in M^o(t')$ for some $t' \in I_i$.  Let $y = \sigma(p)$ for $p = (u,v,t)$.  It follows that $\ell(p) < \|t - t' \| \leq length(I_i) \leq \sep$,  leading to a contradiction.
Hence, $(M,h,I)$ is decomposable.

Suppose now that $(M,h,I)$ is decomposable with width-parameter $\delta$ 
(cf. Definition~\ref{decompDef}).
Consider a point $p_0= (u_0, v_0, t_0) \in {\cal F}$ and 
let $\sigma(p_0) = y_0$.  
Let $I_1 = [t_0 - \delta, t_0]$ and $I_2 = [t_0, t_0 + \delta]$. Further,
let ${\cal E}_i$ and $C_{I_i}$ be the envelope and contact-set for the
sweeps $(M,h,I_i)$ respectively. Observe that 
$y_0 \in C_{I_i}$ for $i=1,2$.
Let ${_{y_0}}R^i = \{ (y,x,t) \in C_{I_i} \times M \times I_i| y = y_0 \}$.
As $(M,h,I)$ is decomposable with width-parameter $\delta$,
both $(M,h,I_1)$ and $(M,h,I_2)$ are simple, and hence,
$C_{I_i} \subset {\cal E}_i$ for $i=1,2$. Therefore, 
$y_0$ belongs to  ${\cal E}_1$ and ${\cal E}_2$.
By Lemma~\ref{simpleLem}, 
${_{y_0}}R^1$ and ${_{y_0}}R^2$ are both singleton sets.  
Further, ${_{y_0}}R^1 = {_{y_0}}R^2 = \{ (y_0,x,t_0) \}$ for 
$x=S(u_0,v_0) \in \partial M$.  Hence, $\ell(p_0) > \delta$.  Since for all $p \in {\cal F}$,
$\ell(p) > \delta$, we conclude that $\sep \geq \delta >0$.

Suppose that $\sep > 0$.  Since $\ell(p) \geq \sep$ for all 
$p \in \partial pT_I$ 
if follows that all the p-trim curves are elementary.
\hfill $\square$

The above proposition provides a {\em natural} test for decomposability. Further, coupled
with Lemma~\ref{transInterLem}, for a decomposable sweep, the problem of
excising the trim set can be reduced to transversal intersections. However,
note that, the very definition of $\sep$ is {\em post-facto} as it 
relies on the trim structures.
Besides, it is the infimum value of the 
not necessarily continuous function $\ell$ and is difficult to compute. Thus, 
the above test of decomposability is not effective.

One of the key contributions of this paper is a novel geometric 
`invariant' function on the funnel which is computed in closed form and 
serves the following objectives.
\begin{enumerate}
\item Quick/efficient and simple detection of decomposability of sweeps, which occur most often in practice.
\item Generation of trim curves for non-decomposable sweeps.
\item Quantification and detection of singularities on the envelope.
\end{enumerate}

For a point $p=(u,v,t) \in \mathcal{F}$, let $q=\sigma(p)$. 
Recall from subsection~\ref{paramSec} that, $J_{\sigma}(p)=[\sigma_u \sigma_v \sigma_t]$ is of rank 2.
As $det(J_{\sigma} (p))=0$, $\{\sigma_u(p), \sigma_v(p), \sigma_t(p)\}$ are linearly dependent. 
Recall that $\sigma_t(p) = V(p)$ is the velocity of the point $S(u,v)$ at time $t$ (cf. subsection~\ref{paramSec}).
As $S$ is regular, the set $\{\sigma_u(p), \sigma_v(p)\}$ forms a basis for the tangent space to $\partial M(t)$.
Therefore, we must have $\sigma_t(p) = l(p).\sigma_u(p) +m(p). \sigma_v(p) $ where $l$ and $m$ are well-defined (unique) on the funnel and are themselves continuous functions on the funnel.

\begin{defn} \label{thetaDef}
The function $\theta : {\cal F} \to \mathbb{R}$ is defined as follows.
\begin{align}	\label{thetaEq}
 \theta (p)= l(p).f_u(p) +m(p).f_v(p) -f_t(p) 
\end{align}
where $f_u, f_v$ and $f_t$ denote partial derivatives of the function $f$ w.r.t. $u,v$ and $t$ respectively at $p$, and $l$ and $m$ are as defined before.
\end{defn}

Note that, unlike $\ell$, $\theta$ is easily and robustly computable continuous 
function on the funnel.
Now we are ready to state one of the main theorems of this paper.

\begin{thm} \label{thetaDecompLem}
If for all $p \in {\cal F}$, $\theta(p) > 0$, then the sweep is decomposable.  
Further, 
if there exists $p \in {\cal F}$ such that $\theta(p) < 0$, then the sweep is 
non-decomposable.
\end{thm}
The proof is given in Section~\ref{proofSec} which highlights
many other surprisingly strong properties of the function $\theta$.

\begin{defn}
The function $\theta$ partitions the funnel ${\cal F}$ into three sets, viz. 
(i) ${\cal F}^+ := \{ p \in {\cal F} | \theta (p) > 0 \}$, 
(ii) ${\cal F}^- := \{ p \in {\cal F} | \theta (p) < 0 \}$ and (iii) ${\cal F}^0 := \{ p \in {\cal F} | \theta (p) = 0 \}$.  
Further, we define ${C_I}^+ := \sigma({\cal F}^+)$, ${C_I}^- := \sigma ({\cal F}^-)$ and 
${C_I}^0 := \sigma ({\cal F}^0)$. 
\end{defn}

\begin{figure}
 \centering
 \includegraphics[scale=0.7]{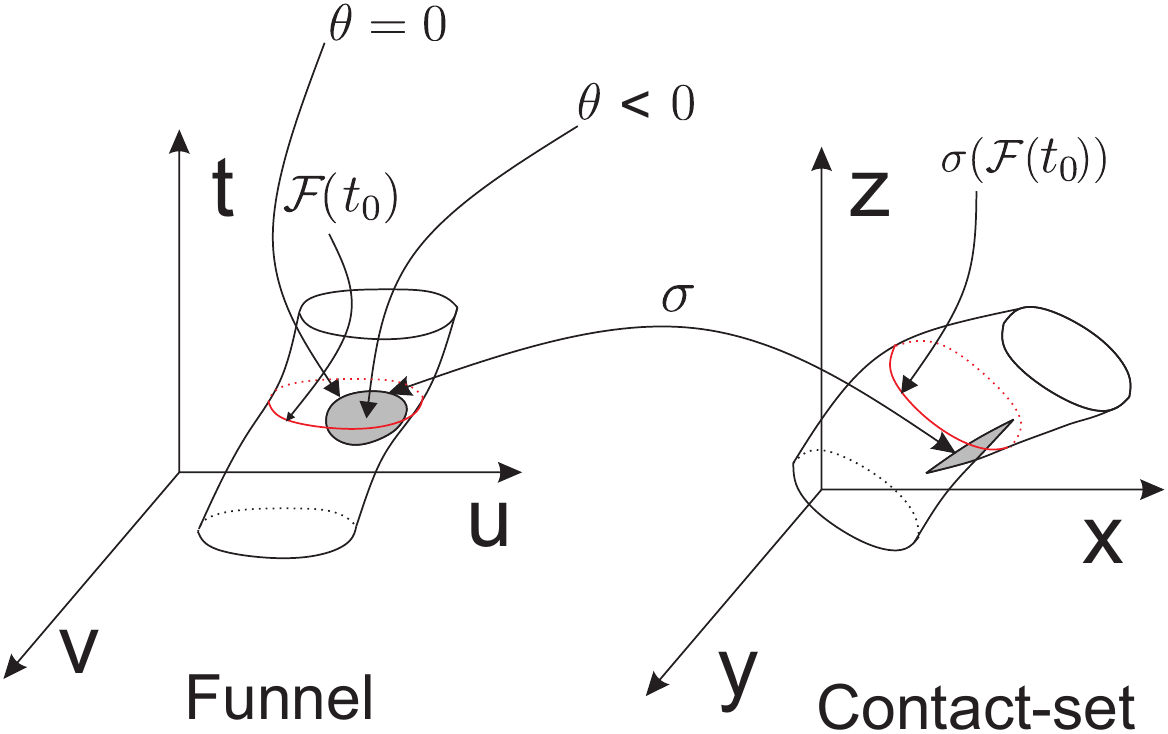}
 \caption{The shaded region on ${\cal F}$ and $C_I$ corresponds to ${\cal F}^-$ and ${C_I}^-$ respectively.  A 
curve of contact is shown in red.}
 \label{lsiRegionFig}
\end{figure}

Figure~\ref{lsiRegionFig} schematically illustrates the sets ${\cal F}^+, {\cal F}^-$ and ${\cal F}^0$ on the funnel 
and sets ${C_I}^-, {C_I}^+$ and ${C_I}^0$.

Note that, for $(M,h,I)$ in general position, 
either ${\cal F}^-$ is a non-empty open set or ${\cal F}={\cal F}^+$. 
Whence, the above theorem provides an efficient `open'
test for decomposability, namely, a sweep $(M,h,I)$
is decomposable iff the open set ${\cal F}^-$ is empty. Most kernels
will have an effective procedure for such a test provided $\theta$ is
effectively computable.

\section{Properties of the invariant $\theta$}  \label{thetaSec}

In this section we prove some key properties of $\theta$, namely, its 
invariance under the re-parametrization of the surface being swept and its relation with the notion of 
inverse trajectory used in earlier works. Finally, we use these properties along
with Proposition~\ref{decompLem}, to
prove Theorem~\ref{thetaDecompLem}.

\subsection{Invariance of $\theta$} \label{thetaInvarSec}
We show that the function $\theta$ is invariant of the parametrization of $\partial M$ and hence, intrinsic to the sweep. 

\begin{thm} \label{thetaInvarThm}
If $\bar{S}$ is a re-parametrization of the surface $S$ so that $q:=\bar{S}(\bar{u}, \bar{v}) = S(u,v)$, and $g(q,t)=0$, then $\theta(u,v,t) = \bar{\theta}(\bar{u}, \bar{v}, t)$.
\end{thm}
\noindent {\em Proof.} 
Suppose as before that the boundary $\partial M$ is specified by the parametrized surface $S$.  Let $\phi: \mathbb{R}^2 \to \mathbb{R}^2$ be a re-parametrization map of $S$ and 
$\bar{S} := S \circ \phi$.  Since $\phi$ is a diffeomorphism, $d\phi$ is an isomorphism at every point in the entire domain of $\phi$.  Let $\phi(\bar{u}, \bar{v}) = (u(\bar{u}, \bar{v}), v(\bar{u}, \bar{v}))$.  
For convenience of expression, we extend $\phi$ to define it on the parameter space of the sweep map $\sigma$ so that $\phi(\bar{u}, \bar{v}, t) = (u,v,t)$.  Hence the re-parametrized 
sweep map (for $\bar{S}$) is simply $\bar{\sigma} = \sigma \circ \phi$.  Recall that $f(u,v,t) = \left < \hat{N}(u,v,t), V(u,v,t) \right>$,
 where $\hat{N}(u,v,t)$ is the unit outward normal to $\partial M(t)$ at 
 the point $A(t)\cdot S(u,v)+b(t)$. It is easy to check that $\hat{N}(u,v,t)$ 
 can also be expressed as $A(t) \cdot (\mathcal{G} \circ S)(u,v)$, where 
$\mathcal{G}:\partial M \to S^2$ is the intrinsic Gauss map, $S^2$ being the unit sphere  
and $\circ$ stands for the usual composition of functions.
Thus, 
\begin{align*}
f(u,v,t) & = \left < \hat{N}(u, v, t), V(u, v, t) \right > \\
& = \left< A(t) \cdot (\mathcal{G} \circ S)(u, v) , V(u,v,t) \right >
\end{align*}

Similarly, computing with the re-parametrization $\bar{S}$, and using
the fact that $\bar{S}=S \circ \phi$, we have $\bar{f}=f \circ \phi$.
\eat{
\begin{align*}
\bar{f}(\bar{u}, \bar{v},t) & = \left < \bar{\hat{N}}(\bar{u}, \bar{v}, t), V(\bar{u}, \bar{v}, t) \right > \\
& = \left< A(t) (\mathcal{G} \circ \bar{S}) (\bar{u}, \bar{v}) , V(\bar{u}, \bar{v}, t) \right >
\end{align*}
}
Differentiating w.r.t. $\bar{u}, \bar{v}$ and $t$ we get
$\nabla \bar{f} = d\phi^T \cdot \nabla f$
where $d\phi$ is the Jacobian of the map $\phi$.
\eat{$d\phi = \begin{bmatrix} u_{\bar{u}} & u_{\bar{v}} & 0 \\ v_{\bar{u}} & v_{\bar{v}} & 0 \\ 0 & 0 & 1 \end{bmatrix}$. } 

Observe that, from Eq.~\ref{thetaEq}, for $\bar{p}=(\bar{u},\bar{v},t)$ and
$p=\phi(\bar{p})=(u,v,t)$, $\theta(p) = \left < \nabla f(p) , z \right >$
where $z = (l, m, -1) $ spans the null-space of $J_{\sigma}|_p$ for $p \in \mathcal{F}$.  In order to compute $\bar{z}$ for the re-parametrized sweep we see that $J_{\bar{\sigma}} = J_{\sigma} \circ d \phi$ and $\bar{z} = d\phi^{-1} z$. 
Now using $\nabla \bar{f} = d\phi^T \cdot \nabla f$,
we get that 
\begin{align*}
\bar{\theta}(\bar{p}) &= \left < \nabla \bar{f}(\bar{p}), \bar{z} \right >  
=\left< d \phi^T \cdot \nabla f(p), d\phi^{-1} \cdot z \right > \\
			      &= \left < \nabla f(p) , z \right > = \theta(p)
\end{align*}
This proves the theorem.
\hfill $\square$

An important corollary of the above theorem is that the function $\theta$
on the funnel is a pull-back of an intrinsic 
function, say $\Theta$, on the abstract smooth manifold
${\cal C}_I= \cup_{t \in I} C_I(t)\times \{t\}$.
More precisely, for $p=(u,v,t) \in {\cal F}$ with $\sigma(p)=y \in C_I(t)$,
define $\Theta((y,t))=\theta(p)$. Then $\Theta$ remains invariant
under a re-parametrization. Observe that, unlike ${\cal C}_I$, in general,
$C_I$ is not a smooth manifold.

\subsection{Geometric meaning of $\theta$} \label{geomThetaSec}
For a smooth point $w$ of $W$, let ${\cal T}_W(w)$ denote the tangent
space to $W$ at $w$.

We show that the function $\theta$ arises out of the relation between two 2-frames on ${\cal T}_{C_I}$. Let $p = (u,v,t) \in \mathcal{F}$ be such that $\sigma(p)$ is a smooth
point of $C_I$. We first compute a natural 2-frame ${\cal X}(p)$ in $\mathcal{T}_{\mathcal{F}}(p)$. 
Note that, $\mathcal{F}$ being the zero level-set of the function $f$, $\nabla f|_p \bot \mathcal{T}_{\mathcal{F}}(p)$.  
We set $\beta := (-f_v, f_u, 0) $ and note that $\beta \bot \nabla f$. It is easy to see that 
$\beta $ is tangent to the p-curve-of-contact $\mathcal{F}(t)$.  
Let $\alpha := \nabla f \times \beta  = (-f_uf_t, -f_vf_t, f_u^2+f_v^2)$.  
Here $\times$ is the cross-product in $\mathbb{R}^3$. Clearly, the set $\{ \alpha, \beta \}$ forms a basis of $\mathcal{T}_{\mathcal{F}}(p)$ if 
$(f_u, f_v) \neq (0,0)$.  Since $\nabla f \neq 0$, if $(f_u, f_v) = (0,0)$ then $f_t \neq 0$ and $\{\alpha', \beta'\} := 
\{(1, 0, 0), (0, f_t, 0) \}$ serves as a basis for $\mathcal{T}_{\mathcal{F}}(p)$.
Figure~\ref{funnelFig} illustrates the basis $\{\alpha, \beta \}$ schematically. 

The set $\{ J_{\sigma} \cdot \alpha , J_{\sigma} \cdot \beta \} \subseteq 
{\cal T}_{C_I}(\sigma(p))$ and can be expressed in terms of 
$\{ \sigma_u, \sigma_v \}$ as follows
\begin{align*}
\begin{bmatrix}
J_{\sigma} \cdot \alpha & J_{\sigma} \cdot \beta
\end{bmatrix}
= 
\begin{bmatrix}
\sigma_u & \sigma_v
\end{bmatrix}
\underbrace{
\begin{bmatrix}
-f_t f_u + l(f_u^2+f_v^2) & -f_v \\
-f_t f_v + m(f_u^2+f_v^2) & f_u
\end{bmatrix} }_{\mathcal{D}(p)}
\end{align*}
Note that,
\begin{align}
det(\mathcal{D}(p)) &= (f_u^2 + f_v^2)(l f_u + m f_v - f_t) 	\label{detDcase1Eq} \\
&=(f_u^2 +f_v^2 ) \theta (p)  \label{thetaDet}
\end{align}
Clearly, if $(f_u, f_v) \neq (0,0)$ then $det(\mathcal{D}(p))$ is a positive scalar multiple of $\theta(p)$.  
Again, if $(f_u, f_v) = (0,0)$, expressing $\{ J_{\sigma} \cdot \alpha', J_{\sigma}\cdot \beta' \}$ in terms of 
$\{ \sigma_u, \sigma_v \}$ we see that $det(\mathcal{D}(p)) = \theta(p) = -f_t$.  

The above relation between $\{ \sigma_u, \sigma_v \}$ and 
$\{ J_{\sigma} \cdot \alpha , J_{\sigma} \cdot \beta \}$ shows that 
if $\theta(p) \neq 0$, then for $y = \sigma(p)$, ${\cal T}_{C_I}(y)$ 
and ${\cal T}_{\partial M(t)}(y)$ are identitical 
(as subspaces of ${\mathbb R}^3$), i.e., 
$\partial M(t)$ makes tangential contact with $C_I$ at $y$.

\subsection{Non-singularity of $\theta$}  \label{thetaNonSingSec}

We give a sweep example which will demonstrate the non-singularity of the function $\theta$.  We show 
that on the set ${\cal F}^0$, $\nabla \theta \neq 0$.
Consider a sphere parametrized as $S(u,v) = (\cos v \cos u, \cos v \sin u, \sin v)$, $v \in [-\frac{\pi}{2}, \frac{\pi}{2}], u \in [-\pi, \pi]$  
swept along a curvilinear trajectory given by $h(t) = (A(t), b(t))$, $A(t) = I, b(t) = (\frac{1}{2} \cos 2t, \frac{1}{2} \sin 2t, 0)$, $t\in [0,1]$.
The unit outward normal at $S(u,v)$ at time $t$ is given by $\hat{N}(u,v,t) = (\cos v \cos u, \cos v \sin u, \sin v)$ and velocity is given by 
$V(u,v,t) = (-\sin 2t, \cos 2t, 0)$.  The envelope function is $f(u,v,t) = \left < \hat{N}(u,v,t) , V(u,v,t) \right > = \cos v \sin (u - 2t)$.  The 
funnel ${\cal F}$ is given by (i) $u = 2t - \pi$, $v \in [-\frac{\pi}{2}, \frac{\pi}{2}]$ and (ii) $u = 2t$, $v \in [-\frac{\pi}{2}, \frac{\pi}{2}]$.  
Hence, $u$ and $v$ can serve as local parameters of ${\cal F}$.  In component (ii) of the funnel, we see that 
 $\theta >0$, hence we will only 
consider component (i).  On ${\cal F}$, $\sigma_t = l \sigma_u + m \sigma_v$ where
$l = \frac{-1}{\cos v}$ and $m = 0$, whence, $\theta(u,v,t) = l f_u + m f_v - f_t = 2 \cos v - 1$.  The set 
${\cal F}^0$ is given by $v = \pm \frac{\pi}{3}$, $u = 2t - \pi$.  
On ${\cal F}^0$, $\frac{\partial \theta}{\partial u} = 0$ 
and $\frac{\partial \theta}{\partial v} = 2 \sin v \neq 0$.   

An important consequence of non-singularity of $\theta$ is that its zero set, i.e., ${\cal F}^0$ 
can be computed robustly and easily.

\subsection{Detecting singularities on the envelope} \label{singSec}

Now we characterize the cusp-singular points of $C_I$. Geometrically, these
are precisely the points where $C_I$ intersects itself non-transversally.
Note that, the transversal singularities of $C_I$ are addressed through
decomposability.
We consider the following restriction of $\sigma$ to the funnel:
$\sigma|_{\cal F}: {\cal F} \rightarrow {\mathbb R}^3$.
Note that $\sigma|_{\cal F}({\cal F})=C_I$.

\begin{defn} \label{singDef}
The set $C_I$ is said to have a {\bf cusp-singularity} at a point $\sigma(p)= x \in C_I$ if 
$\sigma|_{{\cal F}}$ fails to be an immersion at $p$. 
\end{defn}
A basic result about immersion (see \cite{diffTop}) implies that 
if $\sigma|_{\cal F}$ is an immersion at a point $p$, then there is a neighborhood ${\cal N}$ of $p$ such
that $\sigma|_{\cal F}$ is a local diffeomorphism from ${\cal N}$ onto
its image.

\begin{lem} \label{singLem}
Let $p_0 \in {\cal F}$ and $\sigma(p_0) = x_0$. The point  $x_0$ is a 
cusp-singularity iff $\theta(p_0) = 0$.
\end{lem}
\noindent {\em Proof.}  From subsection~\ref{geomThetaSec}, $\theta(p_0)$ is a positive multiple of 
the determinant relating frames $\{ \sigma_u, \sigma_v \}$ and 
$\{ J_{\sigma} \cdot \alpha, J_{\sigma} \cdot \beta \}$ at $x_0$ .  Since the set $\{ \sigma_u, \sigma_v \}$ 
is always linearly independent, it follows that $\{ J_{\sigma} \cdot \alpha, J_{\sigma} \cdot \beta \}$ is linearly 
dependent iff $\sigma|_{{\cal F}}$ fails to be an immersion at $p_0$ iff $\theta(p_0) = 0$.
\hfill $\square$

In other words, the set ${C_I}^0$ is the set of cusp-singular points in $C_I$.

\subsection{Relation with inverse trajectory}  \label{inverseTrajSec}

We now show the relation of the function $\theta$ with inverse 
trajectory~\cite{trimming, classifyPoints} used in earlier works.
Given a trajectory $h$ and a fixed point $x$ in object-space, 
the inverse trajectory of $x$ is
the set of points in the object-space which get mapped to $x$ at some time instant by $h$, i.e. 
$\{ z \in \mathbb{R}^3 | \exists t \in [0,1], A(t) \cdot z + b(t) = x\}$.  

\begin{defn} \label{invTrajDef}
Given a trajectory $h$, the {\bf inverse trajectory} $\bar{h}$ is defined as the map $\bar{h}:I  \to (SO(3), \mathbb{R}^3)$ given by $\bar{h}(t) = (A^t(t), -A^t(t) \cdot b(t))$.  
Thus, for a fixed point $x \in \mathbb{R}^3$, the inverse trajectory of $x$ is the map $\bar{y}:I \to \mathbb{R}^3$ 
given by $\bar{y}(t) = A^t(t) \cdot (x - b(t))$. 
\end{defn}
The range of $\bar{y}$ is $\{ A^t(t) \cdot x - A^t(t) \cdot b(t) | t \in I \} $.  We list some of the facts about 
$\bar{y}$ in the Appendix  which will be used in proving Theorem~\ref{lambdaLem}.

\begin{figure}
 \centering
 \includegraphics[scale=0.7]{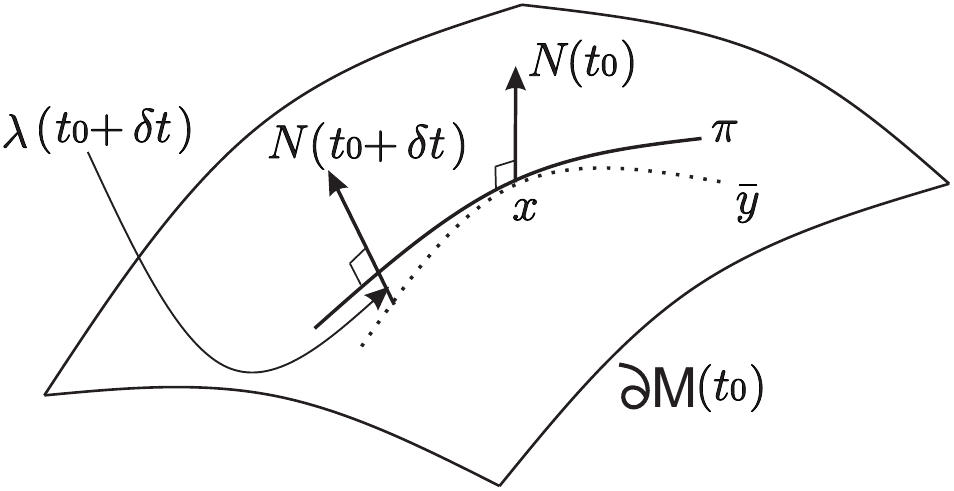}
 \caption{The inverse trajectory of $x$ intersects $M^o(t_0)$.}
 \label{type2LSIFig}
\end{figure}

For the inverse trajectory $\bar{y}$ of a point $x \in \partial M(t_0)$, let $\pi$ be 
the projection of $\bar{y}$ on $\partial M(t_0)$.  Let $\lambda(t)$ be the signed distance 
of $\bar{y}(t)$ from $\partial M(t_0)$. If the point $\bar{y}(t)$ is in $M^o(t_0)$, $Ext(M(t_0))$ 
(the exterior of $M$) or on the surface $\partial M(t_0)$, then $\lambda(t)$ is negative, positive 
or zero respectively.  Then we have $\bar{y}(t) - \pi(t) = \lambda(t)N(t)$, where $\pi(t)$ is the 
projection of $\bar{y}(t)$ on $\partial M(t_0)$ along the unit outward pointing normal 
$N(t)$ to $\partial M(t_0)$ at $\pi(t)$.  This is illustrated in Figure~\ref{type2LSIFig}.
Thus the following relation holds for $\lambda$.
\begin{align} \label{lambdaEq}
\lambda(t) = \left < \bar{y}(t) - \pi(t) , N(t) \right >
\end{align}

\begin{thm} \label{lambdaLem}
For $p= (u_0,v_0,t_0) \in {\cal F}$, 
\begin{align*}
\theta(p) = \ddot{\lambda}(t_0) = \left < -\ddot{\sigma} +2\dot{A} \cdot V , N \right > + \kappa v^2 
\end{align*}
 where $\kappa$ is the normal curvature of $S$ at $(u_0,v_0)$ 
along velocity $V(p)$, $N$ is the unit outward normal to $S$ at $(u_0,v_0)$ and 
$v^2 = \left< V(p), V(p) \right>$.
\end{thm}
See Appendix for the proof.

From Theorem~\ref{lambdaLem}  it is clear that the function $\theta$ is
intimately connected with the curvature of the solid and that of the trajectory. 
It is easy to see that the function $\lambda$ is identical to the function 
$\varphi$ defined in~\cite{trimming} for implicitly defined solids, albeit, is invariant of the function defining the solid 
as well as the parametrization of the same.

\subsection{Proof of Theorem~\ref{thetaDecompLem}}  \label{proofSec}
\noindent {\em Proof.}  Suppose that for all $p \in {\cal F}$, $\theta(p) > 0$.  
For $p \in {\cal F}$, let $t(p)$ denote the $t$-coordinate of $p$.  
Consider the set of points 
$P = \left \{ p \in {\cal F} | \exists p' \in {\cal F}, p' \neq p, \sigma(p) = 
\sigma(p') \right.$ and $\left. \sigma^{-1}(\sigma(p)) =  \{ p, p' \} \right \}$.
By the general position assumption, $P$ is a collection of smooth curves in ${\cal F}$.
For $p \in P$, let $p'$ denote the unique point in $P$ such that $p \neq p'$ and 
$\sigma(p) = \sigma(p')$.
Further, we define $\delta(p) =  \| t(p) - t(p') \|$. 
Let $\delta := \displaystyle \inf_{p \in P} \delta(p)$.
Consider two cases as follows:

{\bf Case (i)}: $\delta = 0$, i.e., there exists a sequence $(p_n)$ in a curve $C$ of $P$ such that
$\displaystyle \lim_{n \to \infty} \delta(p_n) = 0$. Hence there exists $p_0 \in \bar{C}$ 
(closure of $C$)  which is a limit point of $(p_n)$.  Since $\displaystyle \lim_{n \to \infty} \delta(p_n) = 
\displaystyle \lim_{n \to \infty} \| t(p_n) - t(p'_n) \| = 0$ and $\partial M$ is free from self-intersections, 
we have that $\displaystyle \lim_{n \to \infty} \| p_n - p'_n \| = 0$. 
Hence, for a small neighborhood ${\cal N}$ of $p_0$ in ${\cal F}$,
we may parametrize the smooth curve 
$\bar{C} \cap {\cal N}$ by a map $\gamma$ so that $\gamma(0) = p_0$ and, for $s \neq 0$, 
$\gamma(s), \gamma(-s) \in C \cap {\cal N}$ and $\sigma(\gamma(s)) = \sigma(\gamma(-s))$.  
Let $\Gamma(s) := \sigma(\gamma(s))$.  Note that $\Gamma(s) = \Gamma(-s)$. Now, 
\begin{align*}
\frac{d \Gamma}{ds}| _{0} =& \displaystyle \lim_{\Delta s \to 0} \frac{\Gamma(\Delta s) - \Gamma(0)}{\Delta s} 
				= \lim_{\Delta s \to 0} \frac{\Gamma(0) - \Gamma(-\Delta s)}{\Delta s}\\
				=& \lim_{\Delta s \to 0} \frac{\Gamma(0) - \Gamma(\Delta s)}{\Delta s} 
				= \displaystyle -\lim_{\Delta s \to 0} \frac{\Gamma(\Delta s) - \Gamma(0)}{\Delta s} 
\end{align*}

Hence, 
\begin{align*}
\frac{d \Gamma}{ds}\arrowvert _{0} = J_{\sigma}|_{\gamma(0)}. \frac{d \gamma}{ds}|_{0} = 0
\end{align*}
Since $\frac{d \gamma}{ds}|_{0} \in \mathcal{T}_{\mathcal{F}}(p_0)$, 
the map $\sigma|_{{\cal F}}: {\cal F} \to C_I$ fails to be an immersion at $p_0$ 
and by Lemma~\ref{singLem} we get that $\theta(p_0) = 0$, which is a contradiction to the hypothesis.

{\bf Case (ii)}: $\delta > 0$.
Let $\{ I_1, I_2, \ldots, I_k \}$ be a partition of $I$ of width $\frac{\delta}{2}$.  Let ${\cal F}_i$ and ${\cal C}_{I_i}$ 
denote the funnel and the contact set corresponding to subinterval $I_i$.  Then it is clear that for each $i$,
$\sigma: {\cal F}_i \to {\cal C}_{I_i}$ is a diffeomorphism, i.e., for each $i$,
${\cal C}_{I_i}(t) \cap {\cal C}_{I_i}(t') = \emptyset$ for all $t, t' \in I_i$, $t \neq t'$. 
We show that the subproblems $(M,h,I_i)$ are simple for all $i$.  Suppose not, i.e., 
for some $i$, there exists $t \in I_i$ such that $C_{I_i}(t) \cap M^o(t') \neq \emptyset$ for some 
$t' \in I_i$.  Hence the trim set $T_{I_i}$ is not empty.  By Lemma~\ref{transInterLem}, for all but finitely 
many points in 
 $ \partial T_{I_i}$ there are two points $p_1, p_2 \in \partial pT_{I_i}$ such 
that $\sigma(p_1)= \sigma(p_2) = y$.  If $p_1 \in {\cal F}_i(t_1)$ and $p_2 \in {\cal F}_i(t_2)$ then 
it follows that $C_{I_i}(t_1) \cap C_{I_i}(t_2) = y$ leading to contradiction.
Hence, the subproblems $(M,h,I_i)$ are simple for all $i$.  
It follows that $(M,h,I)$ is decomposable with width-parameter $\frac{\delta}{2}$.

Hence we have proved that if for all $p \in {\cal F}$, $\theta(p) > 0$ then the sweep is decomposable.

Suppose now that there exists $p=(u,v,t) \in {\cal F}$ such that $\theta(p) < 0$.  Let $y = \sigma(p)$.  
Recall the definition of the function $\lambda$ from Equation~\ref{lambdaEq} and relation 
$\theta(p) = \ddot{\lambda}(t)$ from Theorem~\ref{lambdaLem}.  Clearly, if $\ddot{\lambda}(t) < 0$,  
then $t$ is a local maxima of the function $\lambda$ and the inverse trajectory of $y$ intersects 
$M^o(t)$.
So, there exists $\epsilon > 0$ such that for all $\delta \in (0, \epsilon)$, 
there exists $w_{\delta} \in M^o(t)$ such that $A(t + \delta) \cdot w_{\delta} + b(t + \delta) = y$.  
Hence, the interval $[t, t+\delta] \subset L(p)$.  Thus $\ell(p) = 0$ and hence $\sep = 0$.
By Proposition~\ref{decompLem}, the sweep is non-decomposable.
\hfill $\square$

\section{Trimming non-decomposable sweeps} \label{nonDecompSec}

\begin{figure}
 \centering
 \includegraphics[scale=0.4]{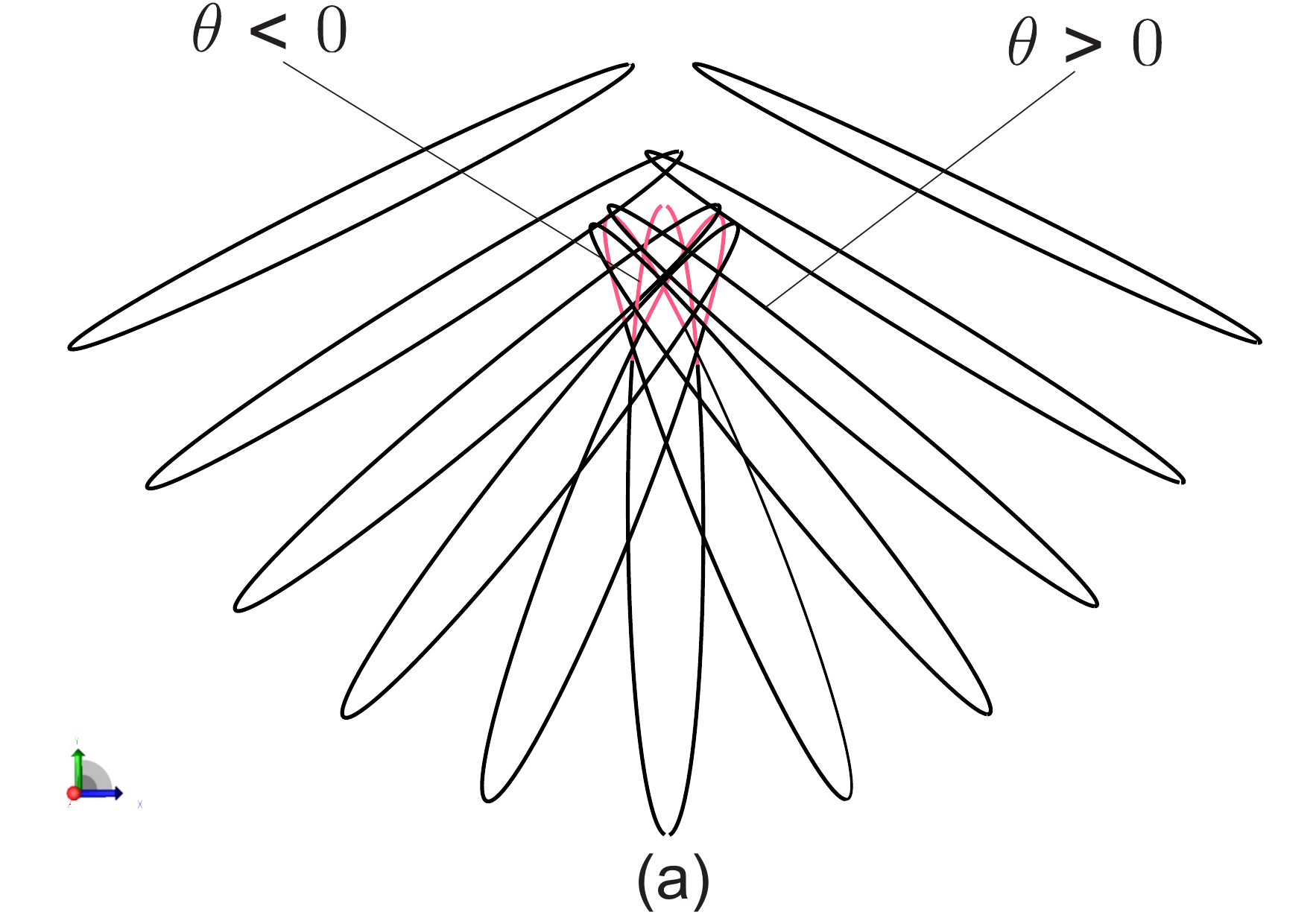}
 \includegraphics[scale=0.4]{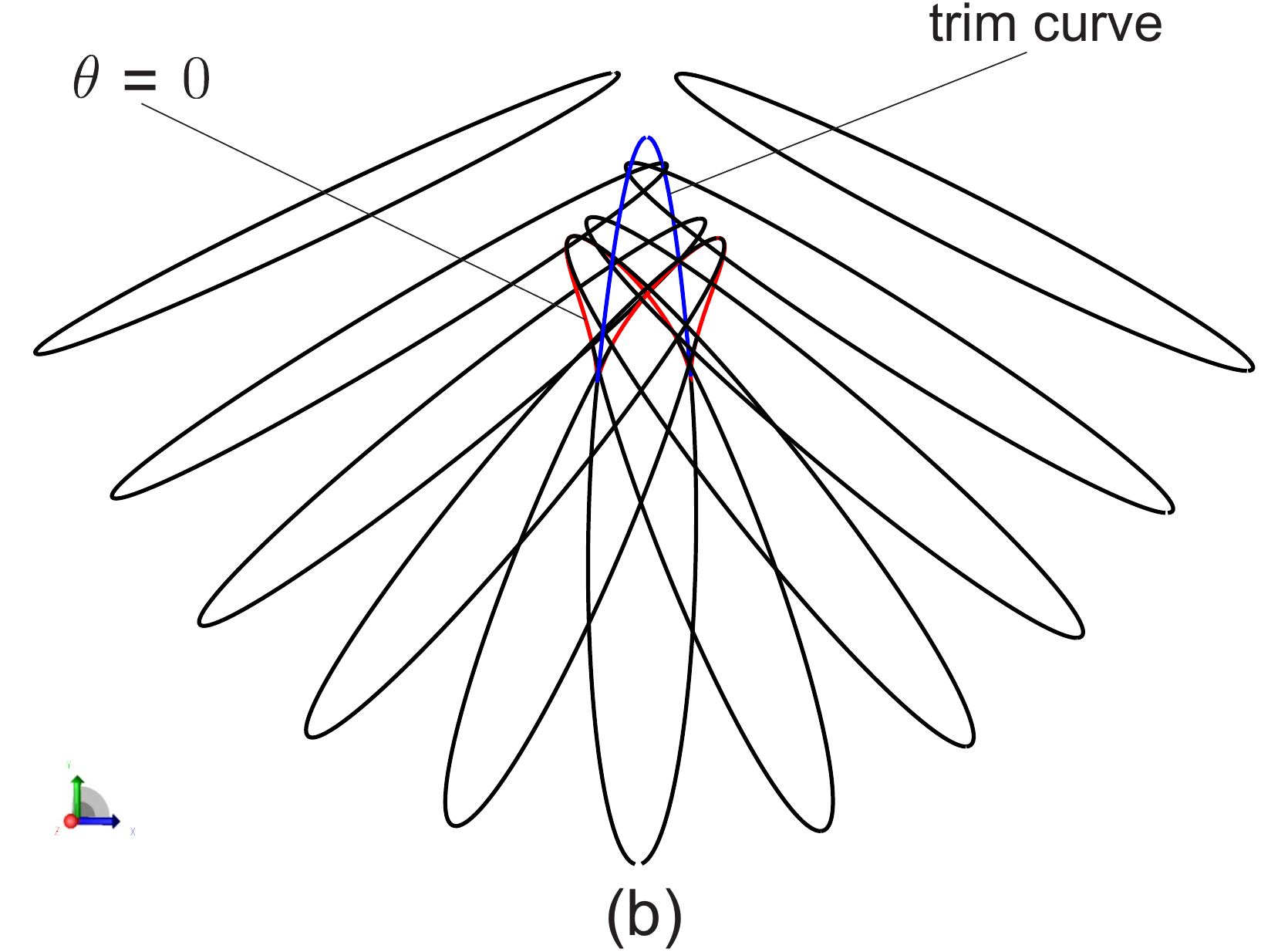}
 \caption{Example of a non-decomposable sweep: a sphere being swept along a parabola (a) Curves of contact at a few time instances (b) The curve $\theta = 0$ is shown in red and trim curve is shown in blue.}
 \label{nonDecompSphereFig}
\end{figure}

\begin{figure}
 \centering
 \includegraphics[scale=0.45]{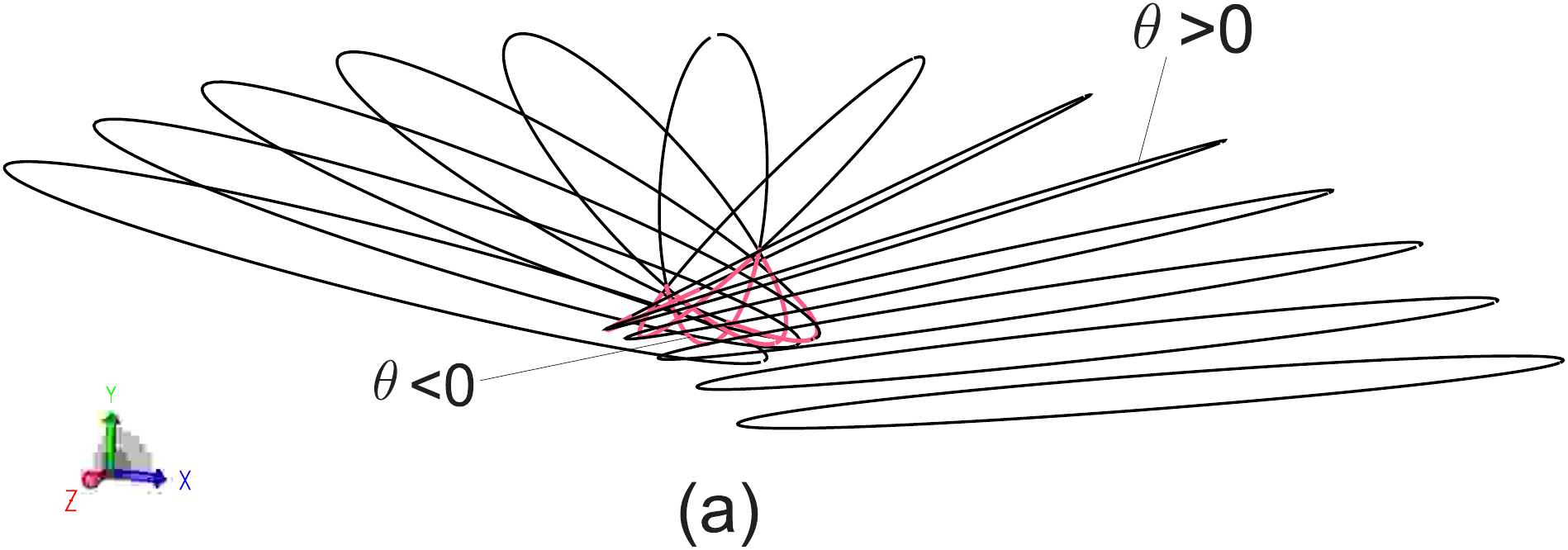}
 \includegraphics[scale=0.45]{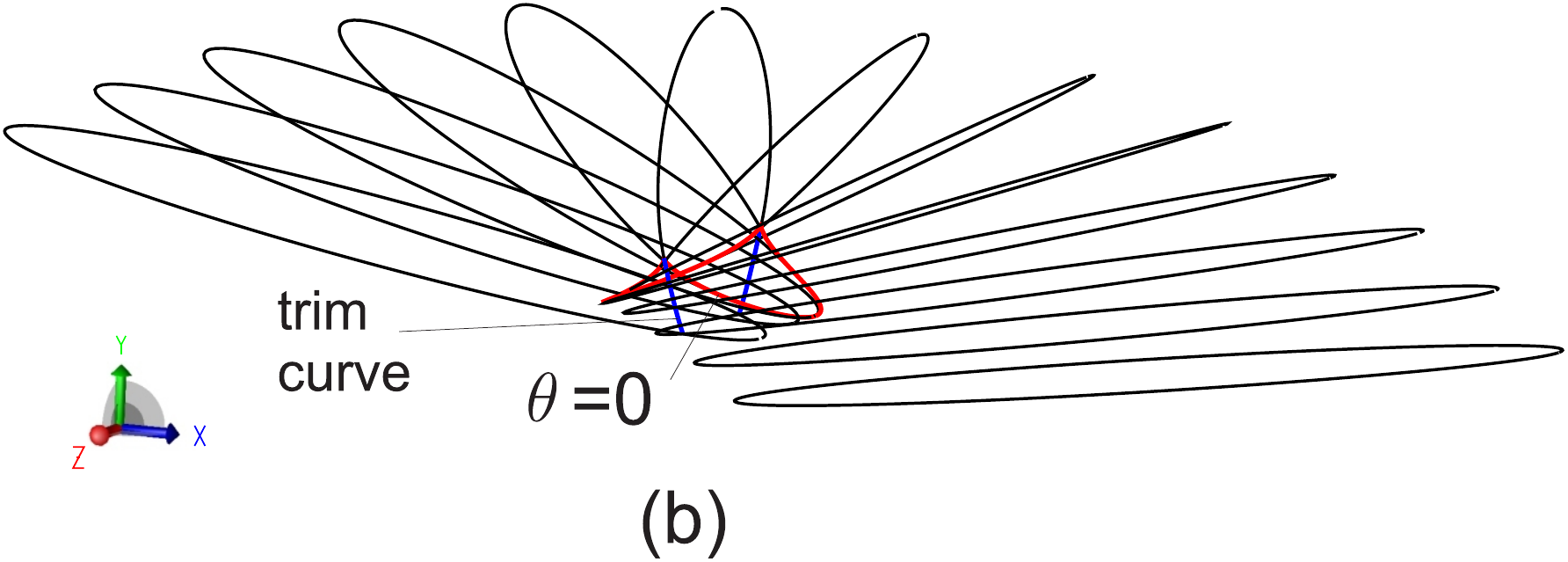}
  \caption{Example of a non-decomposable sweep: an ellipsoid being swept along a circular arc (a) Curves of contact at a few time instances (b) The curve $\theta = 0$ is shown in red and trim curve is shown in blue.}
 \label{nonDecompEllipseFig}
\end{figure}

\begin{figure}
 \centering
 \includegraphics[scale=0.65]{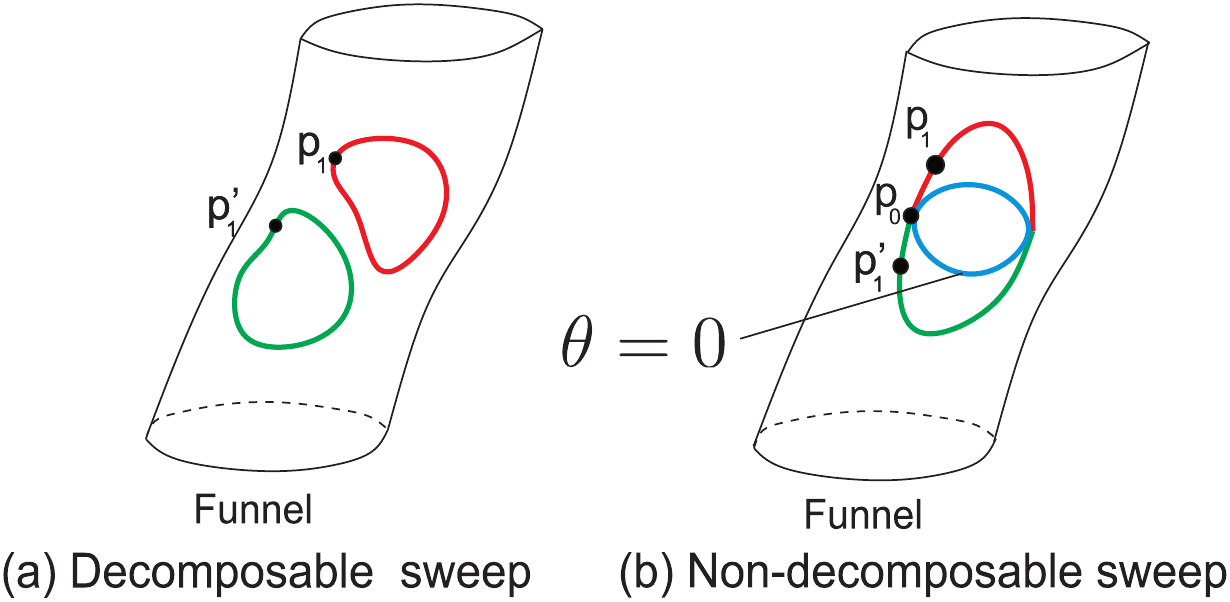}
 \caption{The p-trim curves for decomposable and non-decomposable sweeps shown on ${\cal F}$.  Here, 
$\sigma(p_1) = \sigma(p_1')$. The point $p_0$ is a singular trim point.}
 \label{ptrimCurveFig}
\end{figure}

\begin{figure}
 \centering
  \includegraphics[scale=0.45]{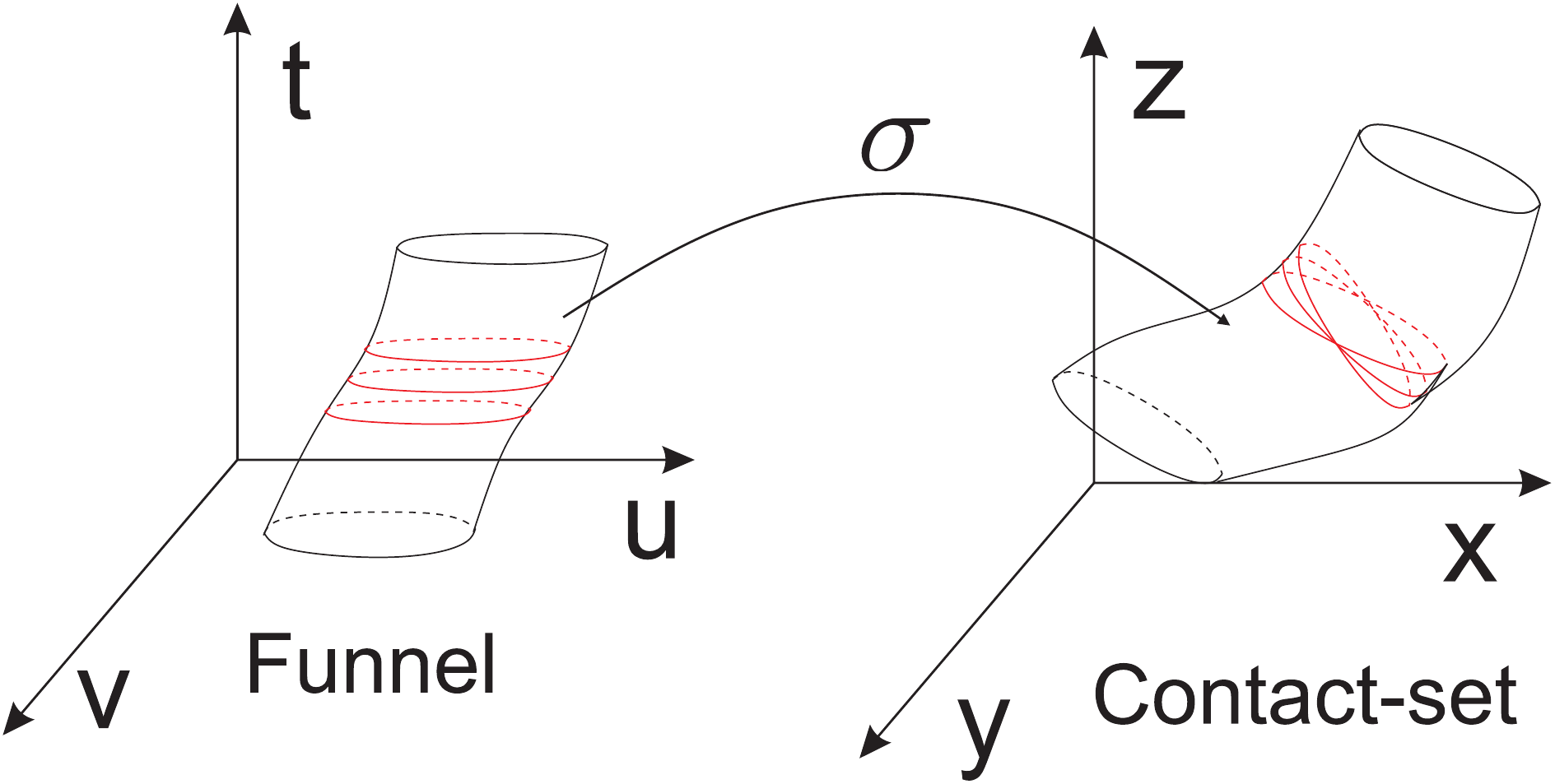}
  \caption{A schematic illustrating the interaction between curves of contact in non-decomposable sweeps.}
 \label{cocFig}
\end{figure}

We recall from Section~\ref{simpleSISec}, the classification
of the curves of $\partial pT_I$ as being elementary or singular. In this
section we look at singular p-trim curves, i.e., a curve $C$ of $\partial pT_I$
where $\displaystyle \inf_{p\in C}  \ell(p) =0$.
Figure~\ref{ptrimCurveFig}(b) schematically illustrates singular p-trim curves.
Figures~\ref{nonDecompSphereFig} and~\ref{nonDecompEllipseFig} show two 
examples of non-decomposable sweeps and the associated singular trim curves.
In Figure~\ref{nonDecompSphereFig} a sphere undergoes curvilinear motion 
along a parabola and in Figure~\ref{nonDecompEllipseFig} an ellipsoid undergoes 
curvilinear motion along a circular arc.  In Figures~\ref{nonDecompSphereFig}(a) 
and~\ref{nonDecompEllipseFig}(a), curves of 
contact at a few time instances are shown. The portions of $C_I(t)$ where $\theta >0$ 
and $\theta < 0$ on ${\cal F}(t)$ are shown in black and pink respectively.  
By Proposition~\ref{thetaNegLem}, the points where $\theta$ is negative do not lie on ${\cal E}$. 
In Figures~\ref{nonDecompSphereFig}(b) and~\ref{nonDecompEllipseFig}(b) such points are excised,  
the curve ${C_I }^0$ is shown in red and the trim curve $\partial T_I$ is 
shown in blue.  Note that ${C_I }^0$ and $\partial T_I$ make contact, which 
they must, as we explain in this 
section.  Figure~\ref{cocFig} schematically illustrates the interaction between curves of contact in 
non-decomposable sweeps.  

\begin{prop} \label{singTrimCLem}
If $C$ is a singular p-trim curve and $p_0 \in  C$ is a limit-point of
$(p_n) \subset C$ such that $\displaystyle \lim_{n \to \infty} \ell (p_n)=0$, then $\theta (p_0)=0$.
\end{prop}
\noindent {\em Proof.}  
The proof is similar to Case (i) of proof for Theorem~\ref{thetaDecompLem}.

\eat{For $p \in {\cal F}$, let $t(p)$ denote the $t$-coordinate of $p$.  By 
Lemma~\ref{trimCLem} it follows that for each $p_n \in (p_n)$ there exists $p'_n \in C$ 
such that $\sigma(p_n) = \sigma(p'_n)$.  Since $\displaystyle \lim_{n \to \infty} \| t(p_n) -t(p'_n) \| = 0$ 
and $\partial M$ is free from self-intersections, we have that $\displaystyle \lim_{n \to \infty} \| p_n - p'_n \| = 0$. 
Hence, for a small neighborhood ${\cal N}$ of $p_0$ in ${\cal F}$,
assuming $C \cap {\cal N}$ is a smooth curve,  we may parametrize 
$C \cap {\cal N}$ by a map $\gamma$ so that $\gamma(0) = p_0$ and 
$\sigma(\gamma(-s)) = \sigma(\gamma(s))$ for $\gamma(-s), \gamma(s) \in {\cal N} \cap C - p_0$.  
Let $\Gamma(s) := \sigma(\gamma(s))$.  Then, 
\begin{align*}
\frac{d \Gamma}{ds}| _{0} =& \displaystyle \lim_{\Delta s \to 0} \frac{\Gamma(\Delta s) - \Gamma(0)}{\Delta s} 
				= \lim_{\Delta s \to 0} \frac{\Gamma(0) - \Gamma(-\Delta s)}{\Delta s}\\
				=& \lim_{\Delta s \to 0} \frac{\Gamma(0) - \Gamma(\Delta s)}{\Delta s} 
				= \displaystyle -\lim_{\Delta s \to 0} \frac{\Gamma(\Delta s) - \Gamma(0)}{\Delta s} 
\end{align*}

Hence, 
\begin{align*}
\frac{d \Gamma}{ds}\arrowvert _{0} = J_{\sigma}|_{\gamma(0)}. \frac{d \gamma}{ds}|_{0} = 0
\end{align*}
Since $\frac{d \gamma}{ds}|_{0} \in \mathcal{T}_{\mathcal{F}}(p_0)$, 
the map $\sigma|_{{\cal F}}: {\cal F} \to C_I$ fails to be an immersion at $p_0$ 
and by Lemma~\ref{singLem} we conclude that $\theta(p_0) = 0$.
\hfill $\square$.
}

\begin{defn} \label{singTrimPtDef}
A limit point $p$ of a singular p-trim curve $C$ such that $\theta(p)=0$ will be called a 
{\bf singular trim point}.
\end{defn}

In Figure~\ref{ptrimCurveFig}(b) a singular trim point $p_0$ is shown on $\partial pT_I$.

\begin{prop} \label{thetaNegLem}
If $p_0 \in {\cal F}$ such that $\theta(p_0) < 0$ then $p_0 \in pT_I$.
\end{prop}
\noindent {\em Proof.}
Let $p_0 = (u_0, v_0, t_0) \in {\cal F}$.
Recall the definition of the function $\lambda$ from Equation~\ref{lambdaEq} and relation 
$\theta(p_0) = \ddot{\lambda}(t_0)$ from Theorem~\ref{lambdaLem}.  Clearly, if $\ddot{\lambda}(t_0) < 0$,  
then $t_0$ is a local maxima of the function $\lambda$ and the inverse trajectory of $\sigma(p_0)$ intersects 
$M^o(t_0)$ and $\sigma(p_0) \in T_I$.  Hence, if $\theta(p_0) < 0$ then $p_0 \in pT_I$.
\hfill $\square$

The above two propositions link the curves of ${\cal F}^0$ to the curves of $\partial pT_I$.
We see that every curve of ${\cal F}^0$ lies inside a curve of $\partial pT_I$ and 
every curve $C$ of $\partial pT_I$ has a curve ${\cal F}^0_C$ of ${\cal F}^0$ 
which makes contact with it.  We have already seen that ${\cal F}^0$ is a collection of curves 
on which $\nabla \theta$ is non-zero.  Thus, the computation of ${\cal F}^0$ in modern 
kernels is straightforward. The task before us is now to locate the points of 
${\cal F}^0 \cap \partial pT_I$. This is enabled by the following function.

\begin{defn}  \label{omegaDef}
Let $\Omega$ be a parametrization of a curve ${\cal F}^0_i$ of ${\cal F}^0$.  
Let $\Omega(s_0) = p_0 \in {\cal F}^0_i$ and  $\bar{z} :=(l,m,-1) \in null(J_{\sigma})$ 
at $p_0$, i.e., $l \sigma_u + m \sigma_v = \sigma_t$.   Define the function 
$\varphi :{\cal F}^0 \to \mathbb{R}$ as follows.
\begin{align}
\varphi(s_0) = \left < \bar{z} \times \frac{d \Omega}{d s}|_{s_0}    , \nabla f|_{p_0} \right >
\end{align}
where $\times$ is the cross-product in $\mathbb{R}^3$.
\end{defn}
 Here, $\varphi$ is a measure of the oriented angle between the tangent at $p_0$ to ${\cal F}^0_i$ 
and the kernel  (line) of the Jacobian $J_{\sigma}$ restricted to the tangent space ${\cal T}_{\cal F}(p_0)$.

\begin{prop} \label{omegaProp}
Every singular p-trim curve $C$ makes contact with a curve ${\cal F}^0_i$ of ${\cal F}^0$ 
so that if
$p_0$ is a singular trim point of $C$ then $\varphi(p_0) = 0$.  Furthermore, at 
such points, $\varphi '(p_0) \neq 0$ where $\varphi '$ refers to the derivative of $\varphi$ 
along the curve ${\cal F}^0_i$.
\end{prop}
\noindent {\em Proof.} We know from Proposition~\ref{thetaNegLem} that ${\cal F}^- \subset pT_I$.  
Since ${\cal F}^0$ and $\partial pT_I$ form the boundaries of ${\cal F}^-$ and $pT_I$ respectively, 
${\cal F}^0$ and a singular p-trim curve $C$ of $\partial pT_I$ meet tangentially at the singular trim point.
Further, by an argument similar to the case (i) of Theorem~\ref{thetaDecompLem}, it can be seen that 
at a singular trim point $p_0$, 
${\cal T}_{C}(p_0)$ is the null-space of the Jacobian $J_{\sigma}|_{p_0}$.  Since 
${\cal T}_{C}(p_0) = {\cal T}_{{\cal F}^0}(p_0)$, $J_{\sigma}|_{p_0}({\cal T}_{{\cal F}^0}(p_0)) = 0$.  
Since the function $\varphi$ measures the oriented angle between $null(J_{\sigma})$ and ${\cal T}_{{\cal F}^0}$, 
it follows that $\varphi(p_0) = 0$.

The derivative $\varphi' \neq 0$ at singular trim points for non-decomposable sweeps shown in 
Figure~\ref{nonDecompSphereFig} and Figure~\ref{nonDecompEllipseFig}.
\hfill $\square$

\eat{
\begin{figure}
 \centering
 \includegraphics[scale=0.5]{plot}
 \caption{The plot of the function $\varphi$ for the sweep example shown in 
Figure~\ref{nonDecompSphereFig}.}
 \label{varphiPlotFig}
\end{figure}
}

Proposition~\ref{omegaProp} confirms that for every singular p-trim curve, we may use the 
function $\varphi$ to locate a singular trim point $p_0$ in a computationally robust manner.  Thus, 
via $\theta$ and $\varphi$ we may access every component of $\partial pT_I$.

\begin{prop}
In the generic situation, (i) the singular p-trim curve $C$ has a finite set of singular trim points.  Each of these 
points lie on a curve of ${\cal F}^0$. (ii) For all but finitely many non-singular points $p \in C$, the image 
$\sigma(p)$ lies on the transversal intersection of two surface patches $\sigma({\cal F}_i)$ 
and the remaining non-singular points lie on intersection of three surface patches $\sigma({\cal F}_i)$
where each ${\cal F}_i \subset {\cal F}$ corresponds to a suitable 
subinterval $I_i \subset I$. 
\end{prop}
\noindent {\em Proof.} 
It follows from Proposition~\ref{singTrimCLem} that the singular trim points lie on ${\cal F}^0$. Since at a 
non-singular trim point $p \in C$, $\ell(p) > 0$, the proof for (ii) is identical to the proof for Lemma~\ref{transInterLem} about elementary trim curves.
\hfill $\square$

Note that the computation of $C$ above is transversal except at the known point $p_0 
\in {\cal F}^0$, i.e., where $\varphi=0$. The problem then reduces to a surface-surface intersection which is transversal except at a known point. This information is usually enough for most kernels to compute $C$ robustly.

\eat{
We now describe the tracing of a singular p-trim curve $C$ once a point $p_0$ in ${\cal F}^0$ has been located 
where $\varphi$ is zero.  
Since ${\cal F}^0$ and $C$ meet tangentially at $p_0$, starting 
at $p_0$ we take small steps in direction $\frac{d \Omega}{ds}|_{p_0}$ and $-\frac{d \Omega}{ds}|_{p_0}$ 
to obtain points $\tilde{p}_1$ and $\tilde{p}'_1$ respectively which are fed to a Newton-Raphson solver which 
returns points $p_1$ and $p'_1$ such that $p_1, p'_1 \in {\cal F}$, $\sigma(p_1) = \sigma(p'_1)$ and 
$t(p_1) - t(p'_1) = t(\tilde{p}_1) - t(\tilde{p}'_1)$.  Let $q_1 := \sigma(p_1) = \sigma(p'_1)$.  
Here, $t(p)$ denotes the $t$-coordinate of $p$.
The point $q_1$ is on the trim curve and the points $p_1$ and $p'_1$ are on the p-trim curve $C$. 
Since points $p_1$ and $p'_1$ are non-singular, these can be fed as starting points to any of the known 
surface-surface intersection algorithms to compute the trim curve. 
}

\begin{figure}
 \centering
 \includegraphics[scale=0.5]{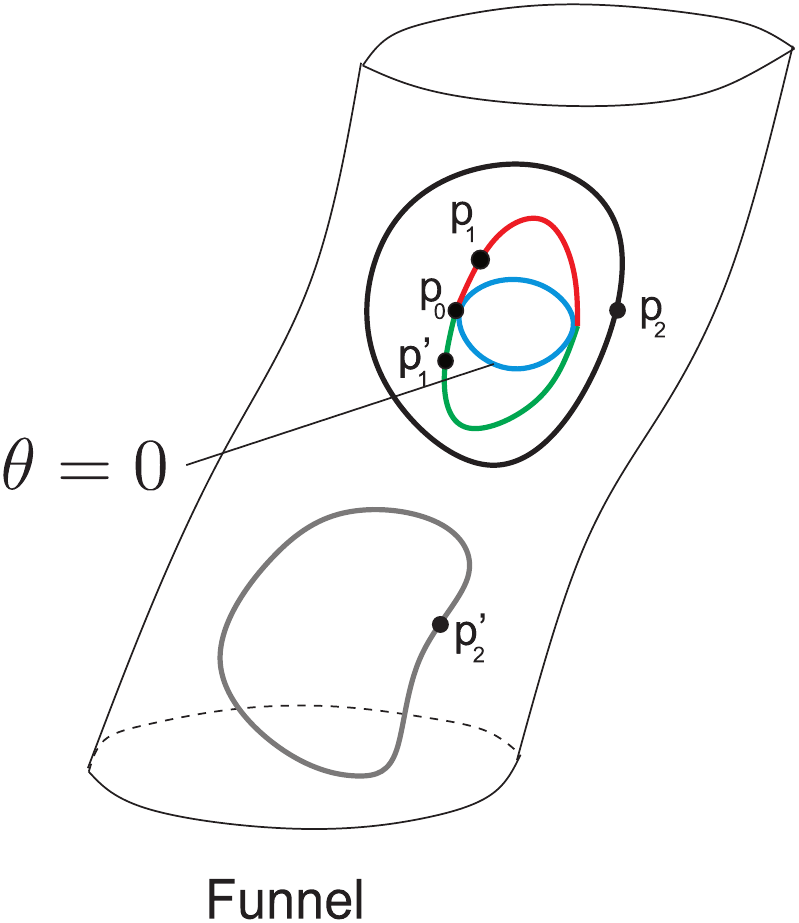}
 \caption{A singular p-trim curve nested inside an elementary p-trim curve}\label{nestedFig}
\end{figure}

Figure~\ref{nestedFig} schematically illustrates a scenario in which a singular p-trim curve is nested 
inside an elementary p-trim curve.  Note that the sweep is non-decomposable and this will be detected 
by the presence of points on ${\cal F}$ where $\theta$ is negative.  Further, the region bounded by the 
singular p-trim curve needs to be excised before a surface-surface intersection algorithm can trace the 
elementary trim curves since no neighborhood of $C_I^0$ (where $\theta$ is zero) can be parametrized. 
Our analysis will first successfully identify and excise the region bound by the singular p-trim curve.  
After parametrizing the remaining part, 
the task of excising the regions bound by elementary p-trim curves can be handled by existing kernels.

\section{Discussion} \label{conclusionSec}

This paper develops a mathematical framework for the implementation of the
``generic'' solid sweep in modern solid modelling kernels. This is done via
a complete understanding of singularities and of self-intersections within
the envelope and the notion of decomposability. This in turn is done
through the
important invariant $\theta $ by which all trim-curves are either stable
surface-surface intersections or are caught by $\theta $.

We now detail certain implementation issues. Firstly, the use of funnel as
the parametrization space and the so called ``procedural'' framework is now
standard, see e.g., the ACIS kernel. Secondly, the non-generic case in the
sweep, as in blends or surface-surface intersections, will need careful
programming and convergence with existing kernel methods for handling
degeneracy. Next, while we have not tackled the case when the trim curves
intersect the left/right caps, that analysis is not difficult and we skip
it for want of space. Finally, the non-smooth sweep is a step away. The
local geometry is already available. The trim curves and other
combinatorial/topological properties of the smooth and non-smooth case are tackled in
a later paper.

Mathematically, our framework may also extend to more complicated cases
where the curves of contact are not simple. This calls for a more
Morse-theoretic analysis which should yield rich structural insights. The
invariant $\theta $ is surprisingly strong and needs to be studied
further.
\eat{
In this paper we give a complete characterization of the trim curves using 
decomposability and the zero-locus of the function $\theta$.  
While the trim curves in decomposable sweeps can be computed by the existing 
surface-surface intersection algorithms, in non-decomposable sweeps 
this approach fails in the neighborhood of singular trim points.  We address this 
problem via the zero-locus of the function $\theta$.
We give examples of simple, decomposable and non-decomposable to illustrate this. 

This work can be extended by computing the complete brep for the envelope which 
includes orienting the edges and faces of ${\cal E}$, computing adjacency relations between them, 
and so on.  The case when faces of $M$ do not meet smoothly needs to be addressed.  
Sweeps in which the number of components of curve of contact changes with time also poses an 
interesting problem.
}
% \appendix

\appendix

\section{Application of solid sweep in design of conveyor screws}

\begin{figure}
 \centering
 \includegraphics[scale=0.6]{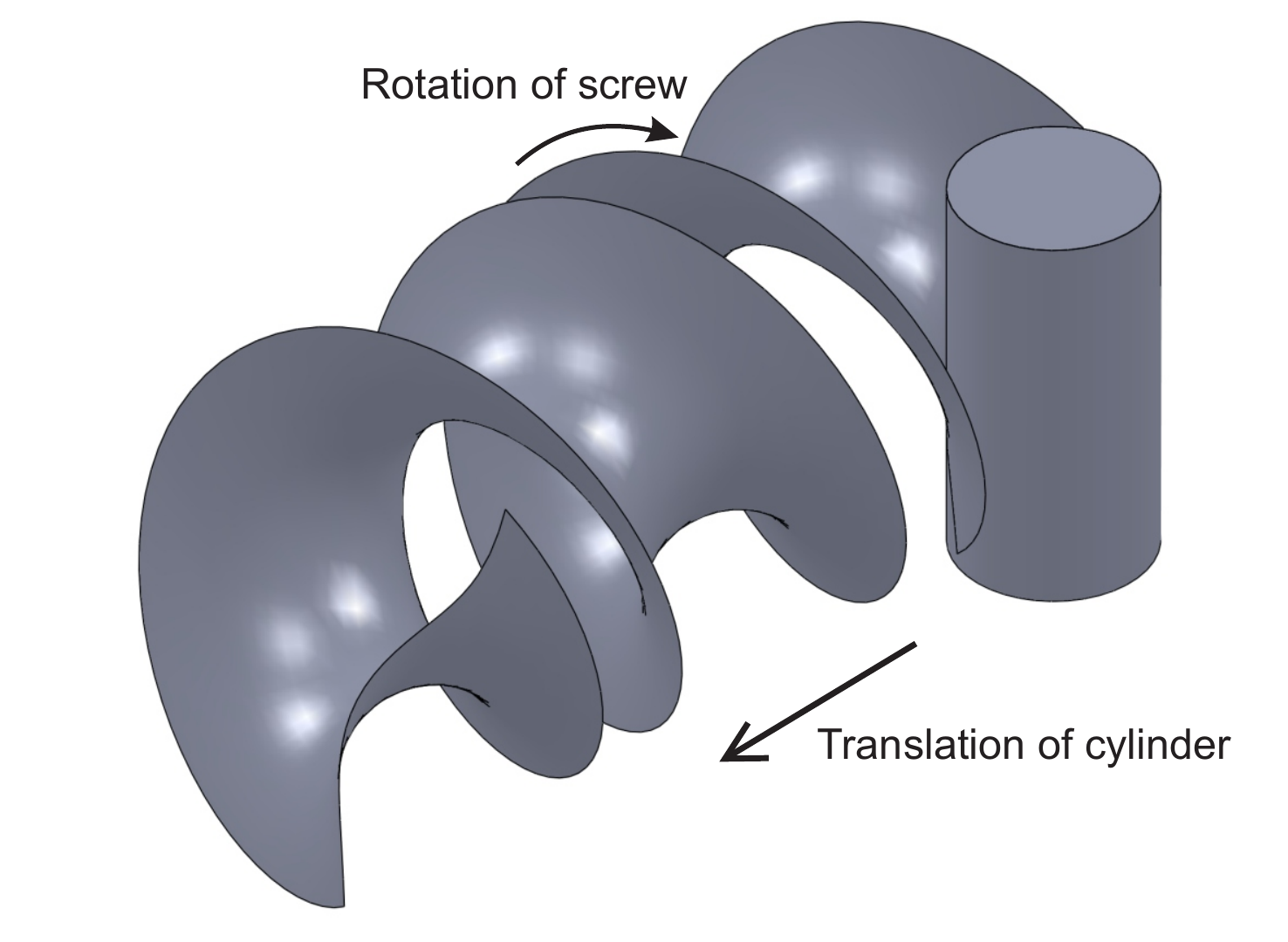}
 \caption{A conveyor screw for translating cylindrical bottles.}
 \label{screwFig}
\end{figure}

\begin{figure}
 \centering
 \includegraphics[angle=0, scale=0.25]{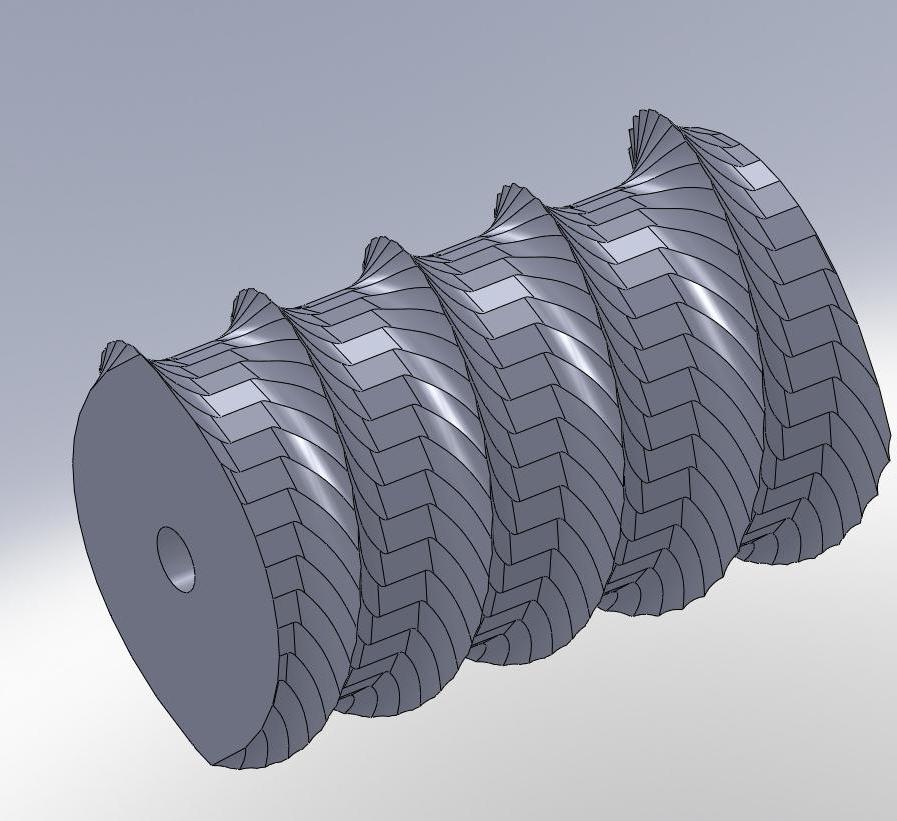}
 \caption{A conveyor screw designed via discrete approach using boolean operations.}
 \label{discreteScrewFig}
\end{figure}

In this section we briefly describe an application of solid sweep in the packaging industry 
where complex needs for handling products arise.  A few example scenarios are, orienting 
the products precisely as they pass along the assembly line, separating one stream of products 
into two streams or combing two streams into one, inverting the product as it passes along the line, 
introducing exact spacing between consecutive products, and so on. 
This is often achieved by a conveyor screw which rotates about its own axis and hence 
propels the product ahead which is sitting in its groove.
The surface of this screw is specifically designed for moving the required product 
along the required path.  See~\cite{scroll} for a video 
of conveyor screws which group a set of products together and introduce precise time lag between 
two consecutive products. 

In order to design such a screw for the required object and the required motion profile, 
the rotation of the screw is compounded into the desired motion profile.  The object is then 
swept along the resultant trajectory and the swept volume so obtained is subtracted from a cylinder to 
obtain the conveyor screw.  Figure~\ref{screwFig} shows the surface of a screw designed 
to translate a cylinder.  The conventional method of designing such screws involves sampling 
the trajectory at a finite number of positions, and taking the union of the object positioned at all these 
positions.  The resultant ``discrete" swept volume is then subtracted from the cylinder to obtain 
an approximate screw.  This is shown in Figure~\ref{discreteScrewFig}.  As expected, this approach 
produces a large number of sliver faces and the brep structure of the resulting solid has a high degree 
of complexity.  Further, the solution is neither accurate nor smooth.

\section{Proof for Proposition~8}	\label{gProofSec}
Recall the statement of Proposition~8 that for $(y, x, t) \in R$ and $I = [t_0, t_1]$, 
either (i) $t = t_0$ and $g(x,t) \leq 0$, or 
(ii) $t = t_1$ and $g(x,t) \geq 0$ or (iii) $g(x,t) = 0$.

\noindent {\em Proof.} 
Define $\hat{e}_1, \hat{e}_2, \hat{e}_3$ and $\hat{e}_4$ as 
\eat{ $\begin{bmatrix} 1 \\ 0 \\ 0 \\ 0 \end{bmatrix}, 
\begin{bmatrix} 0 \\ 1 \\ 0 \\ 0 \end{bmatrix}, \begin{bmatrix} 0 \\ 0 \\ 1 \\ 0 \end{bmatrix}$ and $\begin{bmatrix} 0 \\ 0 \\ 0 \\ 1 \end{bmatrix}$}
$(1, 0, 0, 0), (0, 1, 0, 0),$ $(0, 0, 1, 0)$ and $(0, 0, 0, 1)$ 
 respectively.
We define the following objects in $\mathbb{R}^4$ where the fourth dimension is time.
Let $Z := \{ (A(t) \cdot x + b(t) , t) | \mbox{ where } x \in M \mbox{ and } t \in I \}$ and
$X := \{ (A(t) \cdot x + b(t) , t) | \mbox{ where } x \in \partial M \mbox{ and } t \in I \} $.
Note that $Z$ is a four dimensional topological manifold and $X$ is a three dimensional
submanifold of $Z$. Further, a point $(x,t)$ lies in $Z^o$ if $t \in I^o$ and $x \in M^o(t)$.  
Further, if $I = [t_0, t_1]$, $\partial Z = X \cup (M(t_0), t_0) \cup (M(t_1), t_1)$ 
forms the boundary of $Z$ where Define the 
projection $\mu: \mathbb{R}^3 \times I \to \mathbb{R}^3$ is defined as $\mu(x,t) = x$ and 
the projection $\tau: \mathbb{R}^3 \times I \to \mathbb{R}$ is defined as $\tau(x,t) = t$.
Clearly, for a point 
$w \in \mu(Z)$, if $\mu^{-1}(w) \cap Z^o \neq \emptyset$ then $w \notin {\cal E}$.  Hence a necessary 
condition for $w$ to be in ${\cal E}$ is that the line $\mu^{-1}(w)$ should be tangent to $\partial Z$
which is a three dimensional manifold which is smooth everywhere except at 
$(\partial M(t_0), t_0)$ and at $(\partial M(t_1), t_1)$. 
For $w \in M^o(t_0)$, \eat{ ${\cal T}_{(M^o(0),0)}(w,0)$ is spanned by $\left \{ \hat{e}_1, 
\hat{e}_2, \hat{e}_3 \right \}$.  Hence }
the outward normal to $\partial Z$ at $(w,t_0)$ is given by $-\hat{e}_4$ and  
the outward normal to $\partial Z$ at $(w,t_1) \in (M^0(t_1),t_1)$ is given by $\hat{e}_4$.
We now compute the outward normal to $\partial Z$ at $(w,t) \in X$.
The manifold $X$ is diffeomorphic to $\partial M \times I$, i.e., the cross product of $\partial M$ 
which is a 2-dimensional manifold and $I$ which is a 1-dimensional manifold, with the 
diffeomorphism given by $d: \partial M \times I \to X$, $d(x,t) = (A(t) \cdot x+b(t), t)$.  
Hence, if $\{y_1, y_2 \}$ spans ${\cal T}_{\partial M}(x)$  and $\{1\}$ spans ${\cal T}_{\mathbb{R}}(t)$ 
then the tangent space of $\partial M \times I$ at $(x,t)$ is spanned by 
$\{(y_1, 0), (y_2, 0), \hat{e}_4 \}$ and ${\cal T}_{X}(w,t)$ is spanned by \\
 $\left \{ (A(t) \cdot y_1 , 0),  (A(t) \cdot y_2 , 0),  (A'(t)  \cdot x + b'(t),  1) \right \}$.
Hence, the outward normal to $\partial Z$ at $(w,t)$ is 
$( A(t) \cdot N(x) , \\ - \left < A(t) \cdot N(x) ,v_x(t)  \right >)$.
Consider now three cases as follows.

Case (i): $t=t_0$.  At any point $(w,t_0) \in (\partial M(t_0), t_0)$ there is a cone of outward normals 
given by $\alpha \begin{bmatrix} A(t) \cdot N(x) \\ - \left < A(t) \cdot N(x) , v_x(t)  \right > \end{bmatrix} - \beta \hat{e}_4$ 
where $\alpha, \beta \in \mathbb{R}$ and $\alpha, \beta \geq 0$.  So if the line $\mu^{-1}(w)$ is tangent to $\partial Z$ at $(w,t_0)$ then 
\begin{align*}
\left <  \hat{e}_4,  \alpha \begin{bmatrix} A(t) \cdot N(x) \\ - \left < A(t) \cdot N(x) ,v_x(t)  \right > \end{bmatrix} - \beta \hat{e}_4 \right > = 0
\end{align*}
for some $\alpha, \beta$ where $\alpha > 0$ and $\beta \geq 0$.  Solving the above for $ \left < A(t) \cdot N(x) ,v_x(t)  \right >$ we get $ \left < A(t) \cdot N(x) ,v_x(t)  \right > = - \frac{\beta}{\alpha} \leq 0$.
Hence $g(x,t) \leq 0$.

Case (ii): $t = t_1$.  Proof is similar to case (i).

Case (iii): $t \in I^o$. If the line $\mu^{-1}(w)$ is 
tangent to $X$ at $(w,t)$ there exist $a,b,c \in \mathbb{R}$ not all zero such that 
\begin{align*}
a \begin{bmatrix} A(t) \cdot y_1 \\ 0 \end{bmatrix} + b \begin{bmatrix} A(t) \cdot y_2  \\ 0 \end{bmatrix} + c \begin{bmatrix} A'(t) \cdot x + b'(t) \\ 1 \end{bmatrix} = \hat{e}_4
\end{align*}
It follows that $v_x(t) = A'(t) \cdot x + b'(t) \in span\{ A(t) \cdot y_1,  A(t) \cdot y_2 \} = {\cal T}_{\partial M(t)}(x)$.  
In other words, $\left < A(t) \cdot N(x), v_x(t) \right > = g(x, t) = 0$.
\hfill $\square$

\section{Some useful facts about the inverse trajectory} \label{invTrajSec}

Recall the inverse trajectory of a fixed point $x$  as $\bar{y}(t) = A^t(t) \cdot (x - b(t))$.
We will denote the trajectory of $x$ by $y:[0,1] \to \mathbb{R}^3$, $y(t) = A(t) \cdot x + b(t)$.  We now note a few useful facts about $\bar{y}$. 
We assume without loss of generality that $A(t_0) = I$ and $b(t_0)=0$.  Denoting the derivative with respect to $t$ by $\dot{}$, we have
\begin{align}
\dot{\bar{y}}(t)=\dot{A}^t(t) \cdot (x-b(t)) - A^t(t) \cdot \dot{b}(t)	\label{yBarDotEq}
\end{align}
Since $A \in SO(3)$ we have,
\begin{align}
 A^t(t) \cdot A(t) &= I, \forall t 	\label{aSO3Eq}
\end{align}
Differentiating Eq.~\ref{aSO3Eq} w.r.t. $t$ we get
\begin{align}
\dot{A}^t(t_0) + \dot{A}(t_0) &= 0			\label{aDotTNotEq} \\
\ddot{A}^t(t_0) + 2\dot{A}^t(t_0) \cdot \dot{A}(t_0) + \ddot{A}(t_0) &=0		\label{aDotDotTNotEq}
\end{align}
Using Eq.~\ref{yBarDotEq} and Eq.~\ref{aDotTNotEq} we get
\begin{align}
\dot{\bar{y}}(t_0) &= -\dot{A}(t_0) \cdot x - \dot{b}(t_0) = -\dot{y}(t_0)		\label{yBarDotTNotEq} 
\end{align}
Differentiating Eq.~\ref{yBarDotEq} w.r.t. time we get
\begin{align}
\ddot{\bar{y}}(t) = \ddot{A}^t(t) \cdot (x-b(t)) - 2\dot{A}^t(t) \cdot \dot{b}(t) - A^t(t) \cdot \ddot{b}(t) 	\label{yBarDotDotEq}
\end{align}
Using Equations~\ref{yBarDotDotEq}, \ref{aDotTNotEq} and \ref{aDotDotTNotEq} we get
\begin{align}
 \ddot{\bar{y}}(t_0) = -\ddot{y}(t_0) + 2\dot{A}(t_0) \cdot \dot{y}(t_0)		\label{yBarDotDotTNotEq}
\end{align}

\section{Proof of Theorem 39} \label{proofthm39Sec}
\noindent {\em Proof.}
Recall the definition of function $\lambda$ as 
\begin{align} \label{lambdaEq}
\lambda(t) = \left < \bar{y}(t) - \pi(t) , N(t) \right >
\end{align}
 Differentiating Eq.~\ref{lambdaEq} with respect to time and denoting derivative w.r.t. $t$ by $\dot{}$, we get
\begin{align}
  \dot{ \lambda}(t) &= \left < \dot{\bar y}(t) - \dot{\pi}(t), N(t) \right > + \left < \bar{y}(t) - \pi(t) , \dot{N}(t) \right > \\
\nonumber  \ddot{\lambda}(t) &= \left < \ddot{\bar{y}}(t) - \ddot{\pi}(t), N(t) \right > + 2\left < \dot{\bar y}(t) - \dot{\pi}(t), \dot{N}(t) \right > \\
		&+ \left < \bar{y}(t) - \pi(t), \ddot{N}(t) \right >	\label{ddotLambdaEq}
\end{align}
At $t=t_0$, $\bar{y}(t_0) = \pi(t_0)$.  Since $ \dot{y}(t_0)=V(p) \bot N(p)$, it follows from Eq.~\ref{yBarDotTNotEq}  that $\dot{\bar{y}}(t_0) \bot N(p)$.  It is easy to verify that $\dot{\pi}(t_0) = \dot{\bar{y}}(t_0)$.  Hence, 
\begin{align}
\lambda(t_0) = \dot{\lambda}(t_0) = 0 \label{lambdaTNotEq}
\end{align}
From Eq.~\ref{ddotLambdaEq} and Eq.~\ref{yBarDotDotTNotEq} it follows that
\begin{align}
\nonumber \ddot{\lambda}(t_0) &= \left < \ddot{\bar{y}}(t_0) - \ddot{\pi}(t_0), N(t_0) \right >\\
					&= \left < -\ddot{y}(t_0) + 2\dot{A}(t_0) \cdot \dot{y}(t_0) - \ddot{\pi}(t_0), N(t_0) \right >  \label{ddotLambdaTNotEq}
\end{align}
Since $\pi(t) \in S(t_0)$  for all $t$ in some neighbourhood $U$ of $t_0$, we have that $\left < \dot{\pi}(t), N(t) \right > = 0, \forall t \in U$.  
Hence $\left < \ddot{\pi}(t), N(t) \right> + \left < \dot{\pi}(t), \dot{N}(t) \right > = 0, \forall t \in U$.  
Hence $-\left < \ddot{\pi}(t_0), N(t_0) \right > =  \left< \dot{\pi}(t_0), \dot{N}(t_0) \right > =  \left< \dot{\pi}(t_0), \mathcal{G}^*(\dot{\pi}(t_0)) \right > = \left < \dot{y}(t_0), \mathcal{G}^*(\dot{y}(t_0)) \right>$ = $\left < V(p) , \mathcal{G}^*(V(p)) \right > =\kappa v^2$.  
Here $\mathcal{G}^*$ is the differential of the Gauss map, i.e. the curvature tensor of 
$S(t_0)$ at point $x$.  Using this in Eq.~\ref{ddotLambdaTNotEq} and the fact that  $\dot{y}(t_0) = \dot{\sigma}(p)$, $\ddot{y}(t_0) = \ddot{\sigma}(p)$ we get
\begin{align}
\ddot{\lambda}(t_0)  &= \left < -\ddot{\sigma}(p) + 2\dot{A}(t_0) \cdot V(p) , N(t_0) \right > + \kappa v^2  \label{lsi2Eq}
\end{align}
Recalling that $\theta(p)=lf_u +mf_v -f_t$ 
\begin{align*}
l f_u + m f_v - f_t &= \left< l\hat{N}_u + m\hat{N}_v, V\right> + \left<\hat{N}, l V_u + m V_v \right >\\  &-  \left< \hat{N}_t, V\right > - \left<\hat{N} ,V_t \right >
\end{align*}
Here $\hat{N}_u = \mathcal{G}^*(\sigma_u)$ and $\hat{N}_v = \mathcal{G}^*(\sigma_v)$ 
where $\mathcal{G}^*$ is the shape operator (differential of the Gauss map) of $S(t_0)$ at $(u_0,v_0)$.  
Also, $V_u = A_t \cdot S_u$ and $V_v = A_t \cdot S_v$.  Assume without loss of generality that $A(t_0) = I$ 
and $b(t_0) = 0$, hence $\hat{N} = A(t_0) \cdot N = N$, $\sigma_u = S_u$ and $\sigma_v = S_v$. Using 
Eq.~\ref{aDotTNotEq} and the fact that $V=\sigma_t = l\sigma_u+m\sigma_v$  we get
\begin{align}
\nonumber l f_u + m f_v - f_t &= \left< \mathcal{G}^* \cdot V, V \right > + 2\left<A_t \cdot V ,N \right> - \left <V_t, N\right >  \\
			& = \kappa v^2  + \left < 2A_t \cdot V - V_t  , N \right > \label{lsiRelationEq}
\end{align}
From Eqs.~\ref{lsi2Eq} and~\ref{lsiRelationEq} and the fact that $\frac{\partial \sigma}{\partial t^2}=V_t$ we get 
$\theta(p) = l f_u + m f_v - f_t = \ddot{\lambda}(t_0)$.
\hfill $\square$

\section{Procedural parametrization of the simple sweep} \label{proceduralSec}

We now describe the parametrization of $E:=C_I$ assuming that the sweep $(M,h,I)$ is simple. 
We obtain a procedural parametrization of $E$ which is an abstract way of defining curves and surfaces. 
This approach relies on the fact that from the user's point of view, a parametric surface(curve) in $\mathbb{R}^3$ 
is a map from $\mathbb{R}^2$($\mathbb{R}$) to $\mathbb{R}^3$ and hence is merely a set of programs 
which allow the user to query the key attributes of the surface(curve), e.g. its domain and to evaluate the 
surface(curve) and its derivatives at the given parameter value.  This approach to defining geometry is especially 
useful when closed form formulae are not available for the parametrization map and one must resort to iterative 
numerical methods.  We use the Newton-Raphson(NR) method for this purpose.  As an example, the parametrization 
of the intersection curve of two surfaces is computed procedurally in~\cite{procedural}.  This approach has the 
advantage of being computationally efficient as well as accurate.  For a detailed discussion on the procedural framework, 
see~\cite{sohoni}.

The computational framework is as follows.  Given $S$ and $h$, an approximate funnel is first computed, 
which we will refer to as the seed surface.  Now, when the user wishes to evaluate $E$ or its derivative at some parameter value,  
a NR method will be started with seed obtained from the seed surface.  The NR method will converge, upto the required tolerance, to the required 
point on $E$, or to its derivative, as required.  Here, the precision of the evaluation is only restricted by the finite precision of the computer
 and hence is accurate.  It has the advantage that if a tighter degree of tolerance is required while evaluation of the surface or its derivative, the seed 
surface does not need to be recomputed.  Thus, for the procedural definition of $E$ we need the following:
\begin{enumerate}
\item an NR formulation for computing points on ${E}$ and its derivatives, which we describe in Section~\ref{NRFormSubSec}
\item Seed surface for seeding the NR procedure, which we describe in Section~\ref{seedSubSec}
\end{enumerate}

Recall that by the non-degeneracy assumption, ${E}$ is the union of $E(t) :=C_I(t), \forall t$.  This suggests a natural parametrization of ${E}$ in 
which one of the surface parameters is time $t$.  We will call the other parameter $p$ and denote the seed surface by $\gamma$ which is a map 
from the parameter space of ${E}$ to the parameter space of $\mathcal{\sigma}$, i.e. $\gamma(p,t) = (\bar{u}(p,t), \bar{v}(p,t), t)$ and while 
the point $\sigma(\gamma(p,t))$ may not belong to ${E}$, it is close to ${E}$.  In other words, $\gamma(p,t)$ is close to $\mathcal{F}$.  
We call the image of the seed surface through the sweep map $\sigma$ as the approximate envelope and denote it by $\bar{{E}}$, 
i.e. $\bar{{E}}(p,t) = \sigma(\gamma(p,t))$.  We make the following assumption about $\bar{{E}}$.
\begin{assum} \label{oneOneAssum}
At every point on the \emph{iso-t} curve of $\bar{{E}}$, the normal plane to the \emph{iso-t} curve intersects the \emph{iso-t} curve of ${E}$ in exactly one point.
\end{assum}
Note that this is not a very strong assumption and holds true in practice even with rather sparse sampling of points for the seed surface.  We now describe the Newton-Raphson formulation for evaluating points on ${E}$ and its derivatives at a given parameter value.

\subsection{NR formulation for ${E}$} \label{NRFormSubSec}

Recall that the points on ${E}$ were characterized by the tangency condition $f(u,v,t) = 0$.  
Introducing the parameters $(p,t)$ of ${E}$, we rewrite this equation $\forall (p_0, t_0)$:
\begin{align} \label{envlCondParEq}
\nonumber f(u(p_0,t_0), v(p_0,t_0), t_0 )  &= \left < \hat{N}(u(p_0,t_0), v(p_0,t_0), t_0), \right . \\
			&\left . V(u(p_0,t_0), v(p_0,t_0), t_0) \right > = 0
\end{align}
So, given $(p_0,t_0)$, we have one equation in two unknowns, viz. $u(p_0,t_0)$ and $v(p_0, t_0)$. ${E}(p_0,t_0)$ is defined as the 
intersection of the plane normal to the iso-$t$(for $t=t_0$) curve of $\bar{{E}}$ at $\bar{{E}}(p_0,t_0)$ with the iso-$t$(for $t=t_0$) 
curve of ${E}$ which is nothing but $C_I(t_0)$. Recall that $C_I(t_0)$ is given by $\sigma(u(p, t_0), v(p, t_0), t_0)$ where $u, v, t_0$ obey Eq.~\ref{envlCondParEq}.  
Henceforth, we will suppress the notation that $u,v, \bar{u}$ and $\bar{v}$ are functions of $p$ and $t$.  Also, all the evaluations will be 
understood to be done at parameter values $(p_0,t_0)$.  The tangent to iso-$t$ curve of $\bar{{E}}$ at $(p_0,t_0)$  is given by 
\begin{align}
\frac{\partial \bar{{E}}}{\partial p} =\frac{\partial \sigma}{\partial u} \frac{\partial \bar{u}}{\partial p} + \frac{\partial \sigma}{\partial v} \frac{\partial \bar{v}}{\partial p}
\end{align}
Hence, ${E}(p_0,t_0)$ is the solution of simultaneous system of equations~\ref{envlCondParEq} and~\ref{planeOrthoEq}
\begin{align} \label{planeOrthoEq}
\left < \sigma(u, v ,t_0) - \sigma(\bar{u}, \bar{v}, t_0)   ,   \frac{\partial \bar{{E}}}{\partial p} \right > = 0 
\end{align}
Eq.~\ref{envlCondParEq} and Eq.~\ref{planeOrthoEq} give us a system of two equations in two unknowns, $u$ and $v$ and hence can be put into NR 
framework by computing their first order derivatives w.r.t $u$ and $v$.  For any given parameter value $(p_0,t_0)$, we seed the NR method with the 
point $(\bar{u}(p_0,t_0), \bar{v}(p_0,t_0))$ and solve  Eq.~\ref{envlCondParEq} and Eq.~\ref{planeOrthoEq} for $(u(p_0,t_0), v(p_0,t_0))$ and compute ${E}(p_0,t_0)$.

Having computed ${E}(p,t)$ we now compute first order derivatives of ${E}$ assuming that they exist.  In order to compute $\frac{\partial {E}}{\partial p}$, we differentiate Eq.~\ref{envlCondParEq} and Eq.~\ref{planeOrthoEq} w.r.t. $p$ to obtain
\begin{align} 
&\left < \frac{\partial \hat{N}}{\partial u} \frac{\partial u}{\partial p} + \frac{\partial \hat{N}}{\partial v} \frac{\partial v}{\partial p},  V \right > + \left< \hat{N} , \frac{\partial V}{\partial u} \frac{\partial u}{\partial p} + \frac{\partial V}{\partial v} \frac{\partial v}{\partial p} \right >=0  \label{derPEq1}  \\
\nonumber &\left< \frac{\partial \sigma}{\partial u} \frac{\partial u}{\partial p} +  \frac{\partial \sigma}{\partial v} \frac{\partial v}{\partial p} - \frac{\partial \sigma}{\partial u} \frac{\partial \bar{u}}{\partial p} +  \frac{\partial \sigma}{\partial v} \frac{\partial \bar{v}}{\partial p},   \frac{\partial \bar{{E}}}{\partial p}\right> \\
&+ \left < \sigma(u, v ,t_0) - \sigma(\bar{u}, \bar{v}, t_0) , \frac{\partial ^2 \bar{{E}}}{\partial p^2} \right >= 0   \label{derPEq2}
\end{align}
Eq.~\ref{derPEq1} and Eq.~\ref{derPEq2} give a system of two equations in two unknowns, viz., $\frac{\partial u}{\partial p}$ and 
$\frac{\partial v}{\partial p}$ and can be put into NR framework by computing first order derivatives w.r.t. $\frac{\partial u}{\partial p}$ 
and $\frac{\partial v}{\partial p}$.  Note that Eq.~\ref{derPEq1} and Eq.~\ref{derPEq2} also involve $u$ and $v$ whose computation we have already described.
After computing $\frac{\partial u}{\partial p}$ and $\frac{\partial v}{\partial p}$, $\frac{\partial {E}}{\partial p}$ can be computed as 
$\frac{\partial \sigma}{\partial u} \frac{\partial {u}}{\partial p} + \frac{\partial \sigma}{\partial v} \frac{\partial {v}}{\partial p}$.  $\frac{\partial {E}}{\partial t}$ 
can similarly be computed by differentiating Eq.~\ref{envlCondParEq} and Eq.~\ref{planeOrthoEq} w.r.t. $t$.  Higher order derivatives can be computed in a 
similar manner.

\subsection{Computation of seed surface} \label{seedSubSec}

The seed surface is constructed by sampling a few points on the funnel and fitting a tensor 
product B-spline surface through these points.  For this, we first sample a few time instants, 
say, ${\cal I} =\{t_1, t_2, \ldots, t_n \}$ from the time interval of the sweep.  For each $t_i \in {\cal I}$,  
we sample a few points on the pcurve of contact ${\cal F}(t_i)$.  For this, we begin with one point $p$ on ${\cal F}(t_i)$ 
and compute the tangent to ${\cal F}(t_i)$ at $p$, call it $z$. Then $p+z$ 
is used as a seed in Newton-Raphson method to obtain the next point on ${\cal F}(t_i)$ and this process is repeated.

While we do not know of any structured way of choosing the number of sampled points, in practice even a small 
number of points suffice to ensure that the Assumption~\ref{oneOneAssum} is valid.

\end{document}